\documentclass[reprint,superscriptaddress,amsmath,amssymb,aps]{revtex4-2}
\usepackage{subcaption}
\usepackage{natbib}
\usepackage[colorlinks,linkcolor = blue,citecolor=blue,urlcolor=blue,bookmarks=false,hypertexnames=true]{hyperref}
\usepackage{float}
\usepackage{comment}
\usepackage{mathtools}
\usepackage{graphicx}
\usepackage[export]{adjustbox}
\usepackage{dcolumn}
\usepackage{bm}
\usepackage{sidecap}
\usepackage{xcolor}
\usepackage{tikz}
\begin{document}

\title{Majorana Thermoelectrics and Refrigeration}

\author{Sachiraj Mishra}
\email{sachiraj29mishra@gmail.com}
\affiliation{%
School of Physical Sciences, National Institute of Science Education \& Research, Jatni 752050, India
}%
\affiliation{%
Homi Bhabha National Institute, Training School Complex, Anushaktinagar, Mumbai 400094, India
}%

\author{Ritesh Das}%
\affiliation{%
School of Physical Sciences, National Institute of Science Education \& Research, Jatni 752050, India
}%
\affiliation{%
Homi Bhabha National Institute, Training School Complex, Anushaktinagar, Mumbai 400094, India
}%

\author{Colin Benjamin}%
\email{colin.nano@gmail.com}
\affiliation{%
School of Physical Sciences, National Institute of Science Education \& Research, Jatni 752050, India
}%
\affiliation{%
Homi Bhabha National Institute, Training School Complex, Anushaktinagar, Mumbai 400094, India
}%

\begin{abstract}
A two-terminal quantum spin-Hall heat engine and refrigerator with embedded Majorana bound states (MBS) is analyzed for optimality in thermoelectric performance using the Landaeur-Buttiker approach. This investigation can be an effective tool to detect MBS. Furthermore, the occurrence of MBS can enhance the performance to rival, as well as outperform, some modern nanoscale quantum heat engines and quantum refrigerators. The optimal performance of this MBS quantum heat engine and quantum refrigerator can be further enhanced by an Aharonov-Bohm flux.
\end{abstract}

\maketitle
\section{Introduction}

{``Majorana fermions are quasiparticle excitations that possess non-Abelian statistics and are predicted to exist in certain topological superconductors \cite{KITAEeV20062} and at interfaces where a two-dimensional topological insulator meets an \( s \)-wave superconductor, separated by a ferromagnetic barrier \cite{PhysRevLett.100.096407, PhysRevLett.101.120403}. These fermions have attracted significant interest due to their potential applications in quantum computing and topological quantum technologies. {They are a unique class of fermions with the unusual property of being their own antiparticles.} Their potential applications in quantum computing, particularly in creating qubits resilient to decoherence, have garnered widespread attention \cite{sankar, KITAEeV20062}. This property makes Majorana fermions invaluable in the desing of topological quantum computers. Numerous theoretical proposals have explored the detection of MBS. For instance, by studying electrical conductance, distinctions between topological and trivial superconductors have been identified, providing a promising route for MBS detection \cite{PhysRevLett.98.237002, PhysRevLett.103.237001}. Other techniques such as Hanbury-Brown and Twiss (HBT)-type non-local shot noise correlations \cite{PhysRevB.106.125402} and non-local conductance measurements \cite{PhysRevB.81.085101}, have further served as the toolkit for identifying MBS. Another key feature is the emergence of a \( 4 \pi \)-periodic current-phase relationship in Josephson junctions modeled by superconductor-topological insulator-superconductor systems, a direct consequence of MBS \cite{PhysRevB.79.161408}. Semiconductor-superconductor heterostructures have also been shown to host MBS \cite{Lutchyn2018}. Despite significant theoretical advances, experimental verification of Majorana fermions remains difficult. Observations of zero-bias conductance peak (ZBCP), initially believed to be hallmarks of Majorana zero modes, have often been disputed. For example, claims of MBS detection through ZBCPs in semiconductor nanowires coupled with \( s \)-wave superconductors were later retracted due to experimental inconsistencies \cite{retraction2}. Similar retractions have occurred in studies involving hybrid semiconductor-superconductor systems \cite{gazibegovic2017retracted}. While other works \cite{yin2015observation, Nayak_2021} have also explored ZBCP, these features alone do not conclusively confirm the existence of MBS. Recently, signatures of crossed Andreev reflection calculated in normal metal-superconductor-normal metal junction \cite{PhysRevB.106.125402} and in magnetic topological insulators exhibiting the quantum anomalous Hall effect when in proximity to an \( s \)-wave superconductor like niobium \cite{uday2024induced}. This observation adds a new dimension to the search for MBS, as crossed Andreev reflection provides a robust mechanism for probing these elusive states."} 

Heat engines convert generated heat into practical work, while refrigerators cause heat to flow from a colder terminal to a hotter terminal by utilizing energy or input work. The generation of excess heat in nanoscale devices is a problem that has received considerable attention. This excess heat is either wasted or damages the device. Thermoelectric effects in mesoscopic systems \cite{sothmannoptimal, sothmannog, thermoeffects, benathermo} can be used to design nanoscale heat engines \cite{sothmannoptimal, Whitney, chaoticcavity, engines} that can be used as a conduit to alter the excess heat generated into practical work \cite{thermoeffects, strainedheat}. We propose a setup that can act as a steady-state heat engine \cite{steadystatebetter, strainedheat} or as a steady-state refrigerator \cite{strainedrefg, magnon, threedot} in distinct parameter regimes.

Majorana bound states(MBS) formed due to interacting Majorana fermions can improve thermoelectric performance in the context of a two-terminal quantum spin-Hall heat engine and refrigerator. Thermoelectric devices convert temperature gradients into electrical energy or vice versa using the Seebeck and Peltier effects. The utilization of MBS in such devices can enhance their efficiency.
Furthermore, introducing an Aharonov-Bohm flux can provide additional control over the system's performance. The Aharonov-Bohm effect is a quantum phenomenon that arises when charged particles are influenced by a magnetic field, even when confined to regions with zero magnetic field strength. By manipulating the Aharonov-Bohm flux, one can modulate the behavior of the Majorana bound states and potentially optimize the performance of the quantum heat engine and refrigerator.

This paper studies a two-terminal quantum spin Hall-based Aharonov-Bohm interferometer with embedded Majorana bound states (MBS). We study thermoelectric transport in the presence of MBS by calculating the figure of merit, maximum efficiency, maximum power, and the corresponding efficiency at maximum power for a quantum heat engine and the coefficient of performance (COP) and cooling power for a quantum refrigerator. We show that in the presence of MBS (coupled or individual), the device's performance as a quantum heat engine and quantum refrigerator is significantly improved. We also show that coupling between individual MBS, as well as the coupling between MBS and the interferometer arms, can significantly affect performance. Tuning the Aharonov-Bohm flux can greatly improve the performance of our setup to match or even outperform other contemporary quantum heat engines and refrigerators\cite{strainedheat, engines, strainedrefg, magnon}. For high output power, the corresponding efficiency is low; similarly, for high cooling power, the corresponding coefficient of performance is very low. For our proposed Majorana quantum heat engine/refrigerator to be viable, it must have a high efficiency with finite output power and a high coefficient of performance with finite cooling power. This paper seeks to find such optimal parameter regimes wherein this is possible. There have been several proposals for detecting Majorana bound states with the help of thermoelectric transport \cite{PhysRevB.89.205418, ri2019thermoelectric, chi2020thermoelectric,PhysRevB.105.075418, khim2015thermoelectric, riccotuning, PhysRevB.101.125417, ramosweideman, he2021photon, PhysRevB.105.L081403, zou2022aharonov, weymann2017spin, PhysRevB.95.195425, chen2021high}. There exists another work that provides a detailed analysis of conductance behavior influenced by Majorana interference. This study offers valuable insights into how Majorana bound states affect electronic transport properties, as discussed in \cite{PhysRevB.105.205430}.
The potential application of Majorana fermions in the design of quantum heat engines and quantum refrigerators has been explored rarely, although the figure of merit has been discussed \cite{ramosweideman}. 

The rest of the paper is organized as follows: In Sec. \ref{sec1}, we discuss thermoelectric transport in mesoscopic systems and calculate the various thermoelectric parameters that can be used to quantify the performance of the system as a heat engine or refrigerator using the Onsager relations \cite{beneticasati, sothmannog}. Sec. \ref{sec2} discusses our proposed Aharonov-Bohm interferometer (ABI) and calculates the transmission probability in the presence of MBS using Landauer-Buttiker scattering theory \cite{BUTTIKER1983365, PhysRevB.51.17758}. In Sec. \ref{sec3}, we study the effect of MBS on the setup's performance as a quantum heat engine and a quantum refrigerator. We present our results and analysis in Sec. \ref{sec5} and end with the conclusions in Sec. \ref{sec5}. {The derivation of the formulas for charge and heat currents in our setup is presented in Appendix \ref{App1}.}

\begin{figure*}
    \centering
    \begin{tikzpicture}
        \node (subfig1) at (-2,0) {
            \begin{subfigure}{0.35\textwidth}
                \centering
                \includegraphics[width=\linewidth]{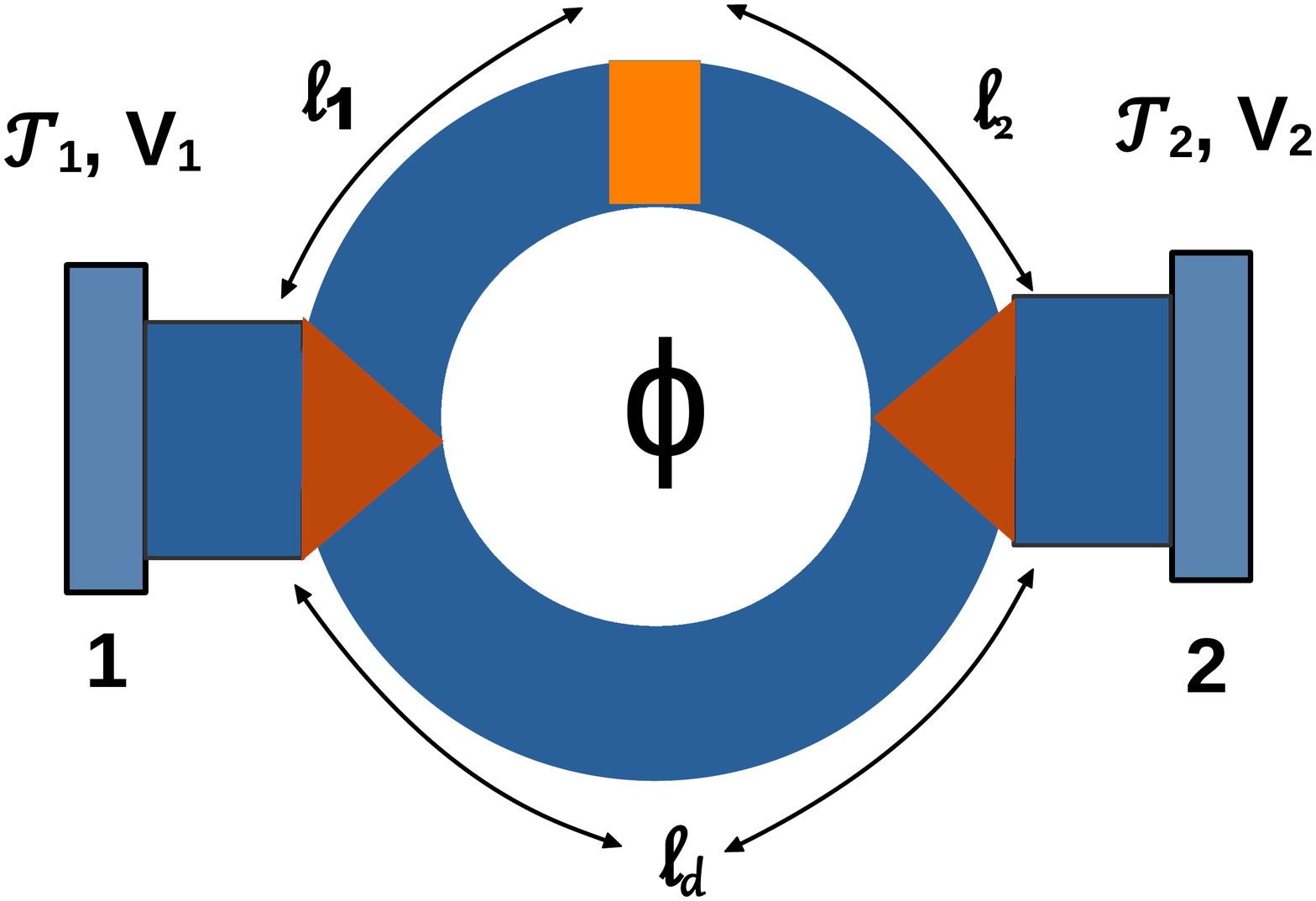}
                \caption{}
            \end{subfigure}
        };

        \node (subfig2) at (6,0) {
            \begin{subfigure}{0.35\textwidth}
                \centering
                \includegraphics[width=\linewidth]{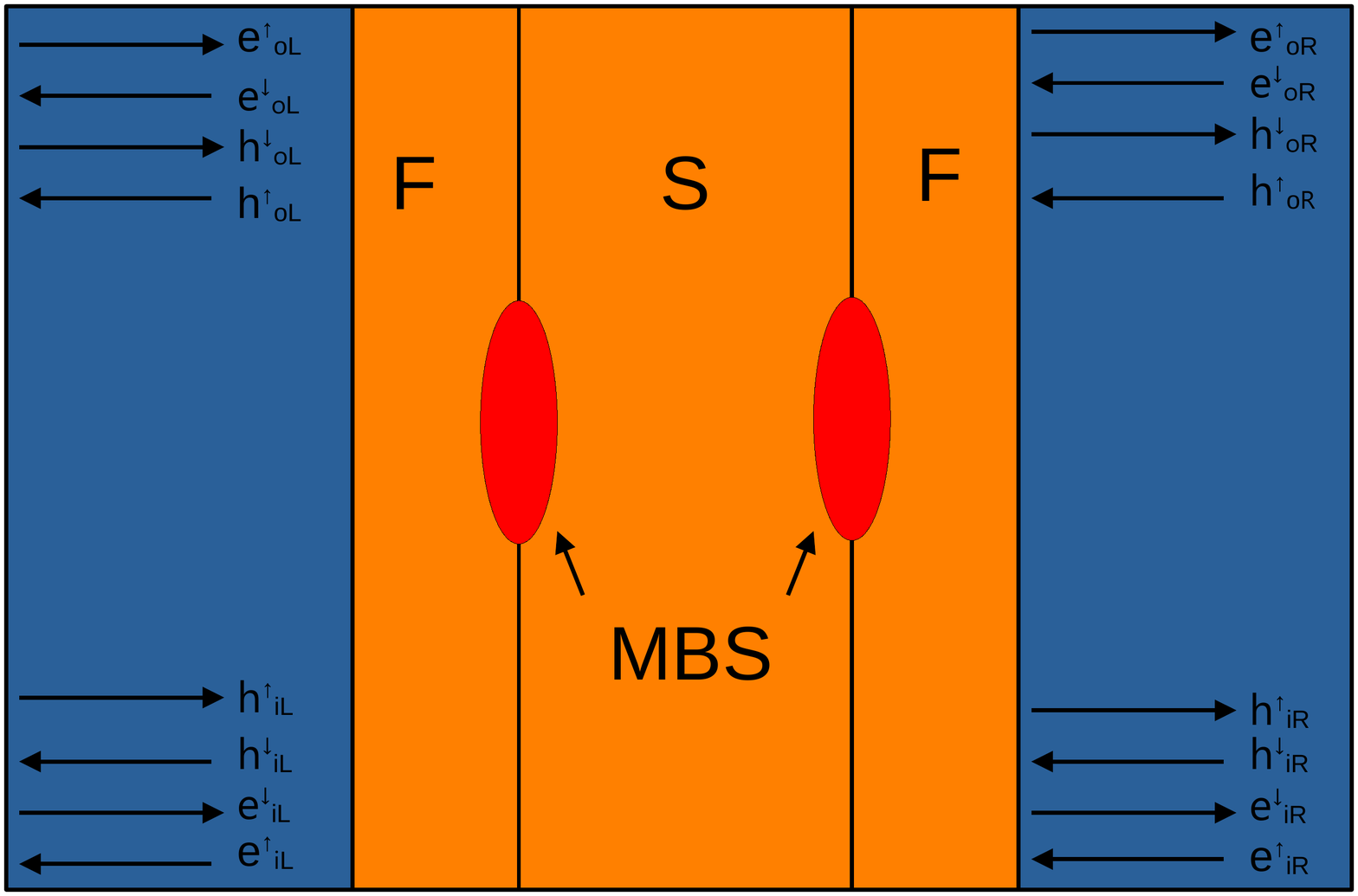}
                \caption{}
            \end{subfigure}
        };

       \node (subfig3) at (2,-8) {
            \begin{subfigure}{0.70\textwidth}
                \centering
                \includegraphics[width=\linewidth]{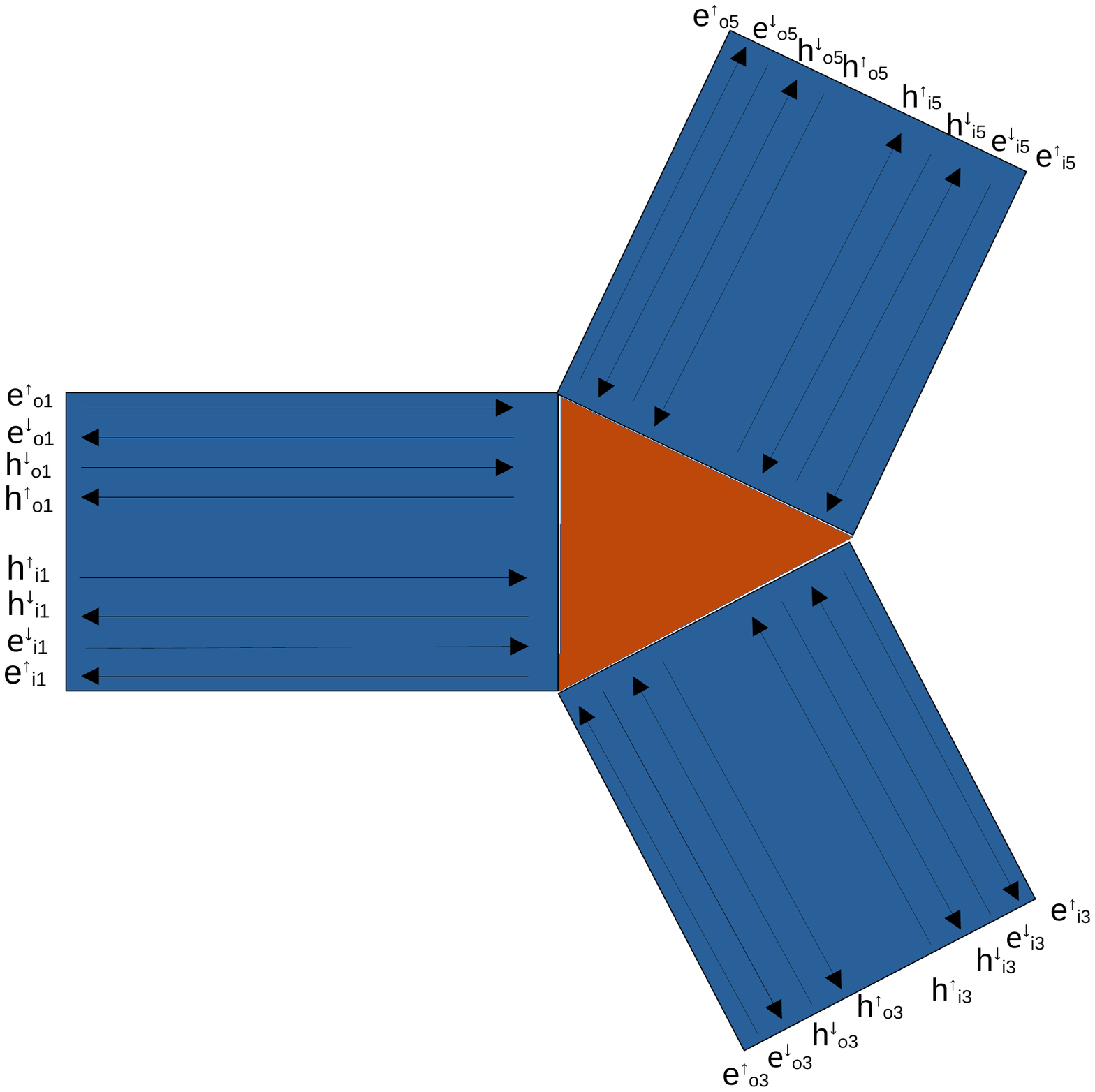}
                \caption{}
            \end{subfigure}
        };          

        \coordinate (start2) at ([xshift=1.65cm, yshift=-0.5cm] subfig1.west);
        \coordinate (end2) at ([xshift=-1cm, yshift=-3cm] subfig3.north);
        \draw[->, thick] (subfig1.north) to[out=30, in=150] (subfig2.north);
        \draw[->, thick] (start2) to[out=-100, in=150] (end2);
    \end{tikzpicture}
    \caption{(a) The proposed MBS-ABI quantum heat engine/quantum refrigerator. It is a 2D TI ring with helical edge modes flowing in the outer and inner edges. The ring is coupled to two leads via couplers (shown in brown). The leads connect the AB ring to reservoirs at temperatures $\mathcal{T}_1$ and $\mathcal{T}_2$, and voltages $V_{1}$ and $V_{2}$. The ring is pierced by an Aharonov-Bohm flux. Ferromagnetic and superconducting (shown in orange) correlations are induced in the top arm of the ring via the proximity effect. (b) The superconductor-topological insulator-ferromagnet (STIM) junction (shown via arrow from (a) to (b)). MBS (shown as red ellipses) occurs at the interface of the ferromagnetic and superconducting (shown in orange) layer. (c) Scattering of the edge modes at the left coupler (shown via arrow from (a) to (c)). $e$ represents electron and $h$ represents hole.}
    \label{fig1}
\end{figure*}

\section{Theory of Thermoelectric Transport in quantum spin hall setups}\label{sec1}
 The total current vector $\textbf{I} = (I, J)$ as derived in Appendix \ref{App1} can be related to force vector $F = (-\Delta V, \Delta \mathcal{T})$ by the Onsager matrix $L$ such that $I = LF$, where \cite{lambert1993multi, PhysRevB.53.16390},
\begin{equation}\label{eq12}
\resizebox{0.9\hsize}{!}{$
L = \begin{pmatrix}
L_{cV} & L_{cT}\\
L_{qV} & L_{qT}
\end{pmatrix}
= \dfrac{4}{h}\int^{\infty}_{-\infty} dE T(E, E_{F})
\xi(E)L_{0}(E, E_{F}),$}
\end{equation}

\begin{equation}\label{eq13}
\resizebox{0.9\hsize}{!}{$
\text{ with $L_{0}(E, E_{F}) = G_{0}$}
\begin{pmatrix}
1 & (E - E_{F})/e\mathcal{T}\\
(E-E_{F})/e & (E-E_{F})^{2}/e^{2}\mathcal{T}
\end{pmatrix},
$}
\end{equation}
 and $f(E) = \left(1 + e^{\frac{E - E_F}{k_B \mathcal{T}}}\right)^{-1}$ being the equilibrium Fermi-Dirac distribution and $\xi(E) =\dfrac{-\partial f(E)}{\partial E},$
where, $G_{0} = (e^{2}/h)$ is the conductance quantum. 
{$T (E, E_F) = \left(2 T_{1  1}^{h  e, \downarrow  \uparrow} + T_{2  1}^{e  e, \uparrow  \uparrow} + T_{2  1}^{h  e, \downarrow  \uparrow} \right)$ is the transmission probability through the setup, where $T_{11}^{he, \downarrow \uparrow}$ is the normal Andreev reflection probability, $T_{21}^{ee, \uparrow \uparrow}$ is the elastic cotunneling probability and $T_{21}^{he, \downarrow \uparrow}$ is the crossed Andreev reflection probability.} From Eq. (\ref{eq12}), we can write \cite{beneticasati}:
\begin{equation}\label{eq15}
I = -L_{cV}\Delta V + L_{cT}\Delta \mathcal{T}
\text{ and } J = -L_{qV}\Delta V + L_{qT}\Delta \mathcal{T}.
\end{equation}
The second law of thermodynamics states that the total entropy must increase in any thermodynamic process. Since the total entropy increases, the entropy production rate $\dot{\xi}$ given by,
\begin{equation}\label{eq16}
\dot{\xi} = -I \Delta V + J\Delta \mathcal{T},
\end{equation}
must be positive. The positivity of entropy production \cite{beneticasati} rate implies

\begin{equation}\label{eq17}
\begin{split}
L_{cV} \geq 0, L_{qT} \geq 0,
L_{cV}L_{qT} - \dfrac{1}{4}(L_{cT} + L_{qV})^{2} \geq 0.
\end{split}
\end{equation}
The output power \cite{beneticasati} is given by,
\begin{equation}\label{eq18}
P = I (V_2 - V_1) = (-L_{cV}\Delta V + L_{cT}\Delta \mathcal{T})\Delta V\, .
\end{equation}
{We consider {$\Delta \mathcal{T} > 0$}. The applied voltage difference $-\Delta V$ and the temperature bias $\Delta \mathcal{T}$ work to operationalize the heat engine [11]. The resulting charge current $I$ given in Eq. (\ref{eq12}) must be such that the output charge power $P$ given in Eq. (12) is positive, which is the condition for the system to perform as a heat engine.} We can maximize the output power by finding $\Delta V$ such that $\dfrac{dP}{d\Delta V} = 0$. For maximum power, we find the voltage bias
\begin{equation}\label{eq19}
\Delta V = \dfrac{L_{cT}}{2L_{cV}}\Delta \mathcal{T}.
\end{equation}
Substituting $\Delta V$ from Eq. (\ref{eq19}) in Eq. (\ref{eq18}), we obtain the maximum power
\begin{equation}\label{eq20}
P_{max} = \dfrac{1}{4}\dfrac{(L_{cT})^2(\Delta \mathcal{T})^{2}}{L_{cV}}\, ,
\end{equation}
and the corresponding efficiency at maximum power \cite{beneticasati} is given by,
\begin{equation}\label{eq21}
\eta_{maxP} = \dfrac{P_{max}}{J} = \dfrac{\eta_{c}\mathcal{T}}{2}\dfrac{L_{cT}^2}{2L_{cV}L_{qT} - L_{cT}L_{qV}}\, ,
\end{equation}
From Eqs. (\ref{eq15}, \ref{eq18}) , we can write the efficiency \cite{beneticasati} $\eta$ as,
\begin{equation}\label{eq22}
\eta = \dfrac{P}{J} = \dfrac{(-L_{cV}\Delta V + L_{cT}\Delta \mathcal{T})\Delta V}{-L_{qV}\Delta V + L_{qT}\Delta \mathcal{T}},
\end{equation}
or
\begin{equation}\label{eq23}
\eta = \dfrac{-(-\Delta V-S \Delta \mathcal{T})\Delta V}{\mathcal{T}S \Delta V - (\dfrac{\kappa}{\sigma}+ \mathcal{T}S^{2})\Delta \mathcal{T})},
\end{equation}
wherein $\sigma = L_{cV}$ is the electrical conductance, $\kappa$ is the thermal conductance \cite{strainedrefg} given by,
\begin{equation}\label{eq24}
\kappa = \dfrac{J}{\Delta \mathcal{T}}|_{I = 0} =\dfrac{L_{cV}L_{qT} - L_{cT}L_{qV}}{L_{cV}},
\end{equation}
and $S = \dfrac{-L_{cT}}{L_{cV}}$ is the Seebeck coefficient \cite{strainedrefg}. Similarly, we can maximize $\eta$ by taking $\Delta V$ such that $\dfrac{d\eta}{d \Delta V} = 0$ in Eq. (\ref{eq22}) with the condition $J > 0$ \cite{beneticasati, strainedheat} to give us maximum efficiency,
\begin{equation}\label{eq25}
\eta_{max} = \eta_{c}\dfrac{\sqrt{Z_T + 1}-1}{\sqrt{Z_T + 1}+1}.
\end{equation}
where $\eta_{c} = \Delta \mathcal{T}/\mathcal{T}$ is the Carnot efficiency and $Z_T$ the figure of merit \cite{beneticasati, strainedheat, ramosweideman} given as,
\begin{equation}\label{eq26}
Z_T = \sigma \dfrac{S^2}{\kappa}e \mathcal{T}\, .
\end{equation}
For the system to work as a quantum refrigerator, net heat current must flow from the cooler to the hotter terminal as net electrical current flows from the higher to the lower potential. The coefficient of performance (COP) is defined as the ratio of cooling power ($P_r$) absorbed to electrical power ($P$) applied and is given as,
\begin{equation}\label{eq27}
COP = \eta^{r} = P_r/{P},
\end{equation}
{where $P_r$ is the heat current extracted from the right terminal 2, i.e., $P_r = -J_2$, i.e., $P_r < 0$, which implies that for the system to work as a refrigerator, {$P < 0$},
maximum COP \cite{strainedrefg} is,}
\begin{equation}\label{eq28}
\eta_{max}^{r} = \eta_{c}^{r}\dfrac{\sqrt{Z_T + 1}-1}{\sqrt{Z_T + 1}+1},
\end{equation}

where $\eta_{c}^{r} = \mathcal{T}/\Delta \mathcal{T}$. The corresponding maximum cooling power at the maximum coefficient of performance is then given as\cite{strainedrefg}
\begin{equation}\label{eq29}
P_{r}^{max} =\Delta \mathcal{T} \sqrt{\dfrac{L_{qT}(L_{cV}L_{qT} - L_{cT}L_{qV})}{L_{cV}}}.
\end{equation}
 The following section discusses the proposed setup and calculates the scattering amplitudes to find the transmission probability $\mathcal{T}(E,E_F) = 2 T_{1   1}^{h   e, \downarrow   \uparrow} +T_{2  1}^{e  e, \uparrow  \uparrow} + T_{2  1}^{h   e, \downarrow   \uparrow}$. The transmission probability can be plugged in Eq. (\ref{eq12}) to find the Onsager coefficients, $L_{cV}, L_{qV}, L_{cT}, L_{qT}$ and the rest of the thermoelectric parameters in Eqs. (\ref{eq20}-\ref{eq29}).

\section{Model}\label{sec2}
\subsection{Hamiltonian}
Our proposed quantum heat engine/refrigerator works based on helical edge modes generated via the quantum spin Hall effect in topological insulators (TIs). We mold a 2D TI into an Aharonov-Bohm ring wherein spin-orbit coupling generates protected 1D edge modes. {The proposed model has been already discussed in Refs. \cite{colin, proobing} and here, we discuss it in brief as also shown in Fig. \ref{fig1}.} The ring is pierced by an Aharonov-Bohm flux $\phi$. Couplers connect the ring to the leads{, which have different voltage and temperature biases.} The Dirac equation for electrons and holes in the ring is given as
\begin{equation}
[v p \tau_{z} \sigma_{z} + (- E_{F} + eA/(\hbar c))\tau_{z}]\psi = E\psi\,.
\end{equation}
$\psi$ is a four-component spinor given by $\psi = (\psi_{e \uparrow}, \psi_{e \downarrow},\psi_{h \downarrow},\psi_{h \uparrow})^{T}$, $p = -i \hbar \partial/\partial x $ is the momentum operator, $E_{F}$ is the Fermi energy, $E$ is the incident electron energy, $v_{F}$ is the Fermi velocity, and $A$ is the magnetic vector potential. MBS (shown in red) occur in the upper half of the ring at the junction between the superconducting and ferromagnetic layer in the TI (STIM junction) (see Fig. \ref{fig1}) \cite{FuKaneMBSog, fuKaneMBS, nilssonMBS}. The Hamiltonian for the MBS is \cite{colin, nilssonMBS, mesaros, proobing}
\begin{equation}
H_{M} = -\sigma_{y}E_{M}\, ,
\end{equation}
with $E_{M}$ denoting coupling strength between individual MBS. The STIM junction is connected to the left and right arms of the ring with coupling strengths $\Gamma_{1}$ and $\Gamma_{2}$, respectively. In the next subsection, we outline the scattering via edge modes in the setup and calculate the transmission probability $T (E, E_F)$.

\subsection{Transport in the system via edge modes}
In a simple quantum Hall ring with an Aharonov-Bohm flux, localized flux-sensitive edge modes develop near the hole, while in leads, edge modes are {not} sensitive to flux \cite{colin, proobing}. To tune the device with an Aharonov-Bohm flux, we must couple the edge modes in the leads and the edge modes in the ring to make the net conductance flux-sensitive. It can be achieved via couplers in the system that can couple the inner and outer edge modes via inter-edge scattering and backscattering. The STIM junction at the top of the ring generates MBS and acts as a backscatter, mixing the electron and hole edge modes. Considering the electron and hole edge modes of spins up and down, we get eight edge modes, with four edge modes circulating on the outer edge and four edge modes circulating on the inner edge. Since no spin-flip scattering occurs in our setup, we can significantly simplify the calculation by dividing the edge modes into two sets of edge modes of opposite spin that scatter as mirror images. It allows us to calculate the transmission for a single set and double it to get the net conductance.

The first set consists of the spin-up electron and spin-{down} hole edge modes. The second set consists of a counterpropagating spin-down electron and spin-{up} hole edge modes. The STIM junction couples the incoming and outgoing edge modes of each set. Among the spin-up {electron and spin-down hole} edge modes, incoming edge modes into the STIM junction are given by {$I_{MBS1} = (e^{\uparrow}_{oL}, e^{\uparrow}_{iR}, h^{\downarrow}_{oL}, h^{\downarrow}_{iR})$} and the outgoing edge modes are given by {$O_{MBS1} = (e^{\uparrow}_{oR}, e^{\uparrow}_{iL}, h^{\downarrow}_{oR}, h^{\downarrow}_{iL})$} (see Fig. \ref{fig1} (b)). Similarly, among the spin-down {electron and spin-up hole} edge modes, the incoming edge modes are {$I_{MBS2} = (e^{\downarrow}_{oR}, e^{\downarrow}_{iL}, h^{\uparrow}_{oR}, h^{\uparrow}_{iL})$}. In contrast, the outgoing edge modes are {$O_{MBS2} = (e^{\downarrow}_{oL}, e^{\downarrow}_{iR}, h^{\uparrow}_{oL}, h^{\uparrow}_{iR})$}. The scattering in each case is the exact mirror image of the other. We can relate the incoming and outgoing edge modes using a $4 \times 4$ scattering matrix $S_{MBS}$ such that $O_{MBSi} = S_{MBS}I_{MBSi}, i \in \{1,2\}$, where $S_{MBS}$ is given by \cite{nilssonMBS, colin, proobing}

\begin{subequations}
\begin{equation}
S_{MBS} = \begin{pmatrix}
1 + ix & -y & ix & -y\\
y & 1 + ix' & y & ix'\\
ix & -y & 1 + ix & -y\\
y & ix' & y & 1 + ix'
\end{pmatrix}
\end{equation}
where
\begin{equation}
\begin{split}
x =\dfrac{\Gamma_{1}(E + i\Gamma_{2})}{z},x' = \dfrac{\Gamma_{2}(E + i\Gamma_{1})}{z},\\
y = \dfrac{E_{M}\sqrt{\Gamma_{1}\Gamma_{2}}}{z},
z=E_{M}^{2}-(E + i\Gamma_{1})(E + i\Gamma_{2})\, ,
\end{split}
\end{equation}
\end{subequations}

The couplers (see Fig. \ref{fig1} (c)) couple the inner and outer edge modes via backscattering and the ring to the leads. {The couplers couple the incoming spin-up electron edge modes to the outgoing spin-up electron edge modes. Similarly, the incoming spin-down electron edge modes are coupled to the outgoing spin-down electron edge modes. For holes, the incoming spin-up edge modes are coupled to the outgoing spin-up edge modes, and the incoming spin-down edge modes are coupled to the outgoing spin-down edge modes.
Thus, the scattering by the couplers is intra-spin and intra-particle, i.e., edge modes outside the ring are
coupled to edge modes of the same spin and same particle as themselves inside the ring.} In the left coupler, the incoming spin-up {electron and spin-down hole edge modes} are given by {$I_{1} = (e^{\uparrow}_{o1},h^{\downarrow}_{o1},e^{\uparrow}_{o3},h^{\downarrow}_{o3},e^{\uparrow}_{i5},h^{\downarrow}_{i5})$}, and the corresponding outgoing waves are given by {$O_{1} = (e^{\uparrow}_{i1},h^{\downarrow}_{i1},e^{\uparrow}_{i3},h^{\downarrow}_{i3},e^{\uparrow}_{o5},h^{\downarrow}_{o5})$}. Similarly, the incoming spin-down {electron and spin-up hole} edge modes are given by {$I_{2}=(e^{\downarrow}_{i2},h^{\uparrow}_{i2},e^{\downarrow}_{i4},h^{\uparrow}_{i4},e^{\downarrow}_{o6},h^{\uparrow}_{o6})$}, and the corresponding outgoing edge modes are {$O_{2} = (e^{\downarrow}_{o2},h^{\uparrow}_{o2},e^{\downarrow}_{o4},h^{\uparrow}_{o4},e^{\downarrow}_{i6},h^{\uparrow}_{i6})$}. The scattering matrix for the couplers is a $6 \times 6$ matrix $S$ such that $O_{i} = SI_{i}, i \in \{1,2\}$ is given by \cite{colin, buttikerboth, proobing}
\begin{equation}\label{eq1001}
S =
\begin{pmatrix}
-(p+q)I & \sqrt{\epsilon}I & \sqrt{\epsilon}I\\
\sqrt{\epsilon}I & pI & qI\\
\sqrt{\epsilon}I & qI & pI
\end{pmatrix},
\end{equation}

where $p = \dfrac{1}{2}(\sqrt{1 - 2\epsilon} - 1)$ and $q = \dfrac{1}{2}(\sqrt{1 - 2\epsilon} + 1$, $I$ is the $2 \times 2$ identity matrix, and $\epsilon$ is a dimensionless parameter which denotes the coupling between the leads and the ring with $\epsilon = 1/2$ for maximum coupling and $\epsilon = 0$ for a completely disconnected loop. While traversing the ring, the spin-up electrons and holes in the edge modes acquire a propagating phase as follows \cite{colin, proobing}:\\
In the upper arm, left of the STIM junction:
\begin{figure*}[]
\center
\includegraphics[width=0.6\textwidth]{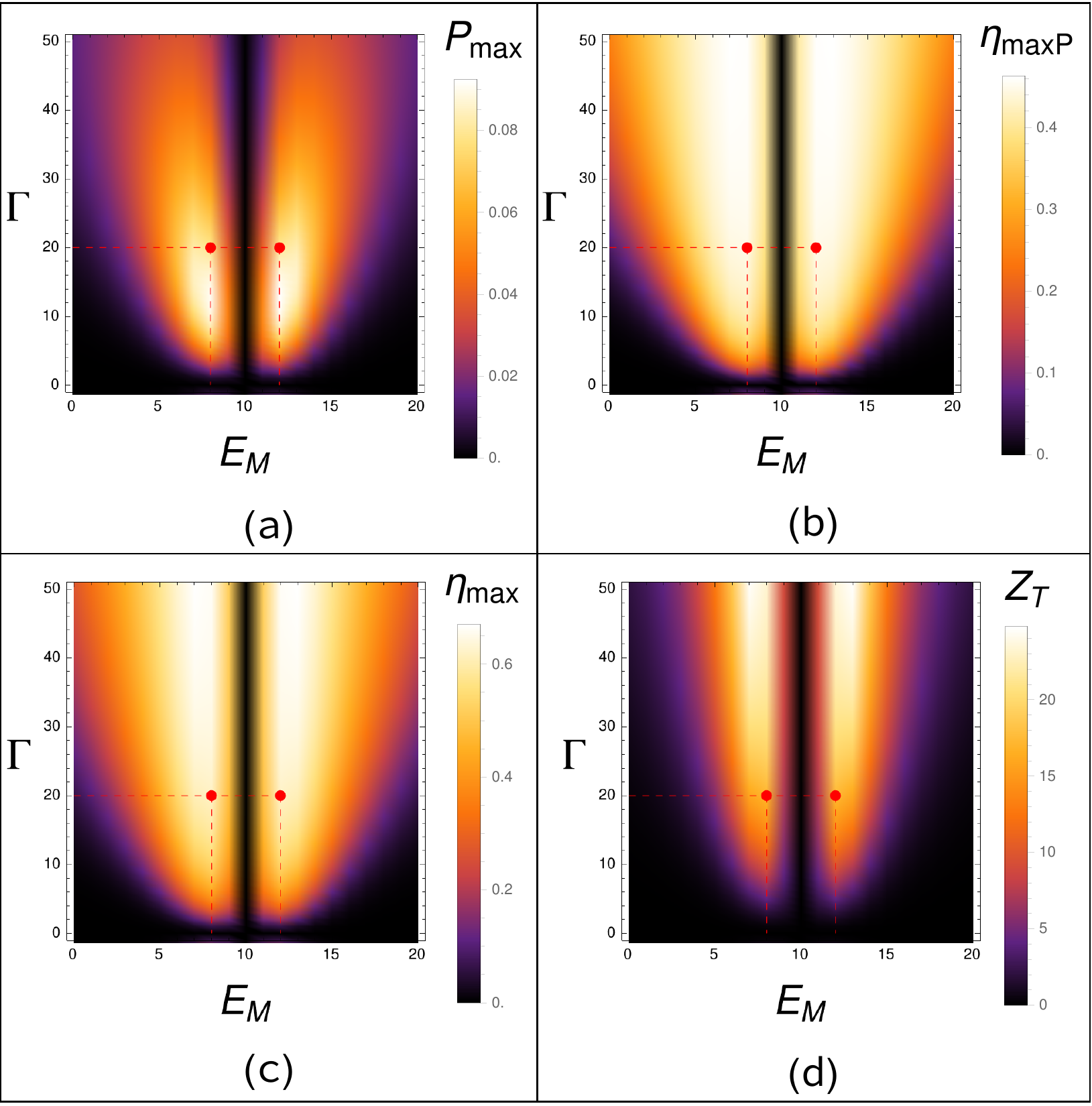}
\caption{Thermoelectric parameters for the Majorana quantum heat engine (a) Maximum power in units of $P_{0} = (k_{B}^{2}T \Delta \mathcal{T})/h $, (b) Efficiency at maximum power $\eta_{maxP}$ in units of $\eta_{c}$, (c) maximum efficiency $\eta_{max}$ in units of $\eta_{c}$, {(d) the figure of merit $Z_T$} vs. $\Gamma$ (in $\mu eV)$ and $E_{M}$ (in $\mu eV)$ for $E_{F} = 10 \mu eV$, $\epsilon = 0.5$, $\phi = \frac{\pi}{2} \phi_{0}$, with $\phi_{0} = hc/e$. The optimal regimes are highlighted with red points.}
\label{fig2}
\end{figure*}

\begin{figure*}[]
\center
\includegraphics[width=0.6\textwidth]{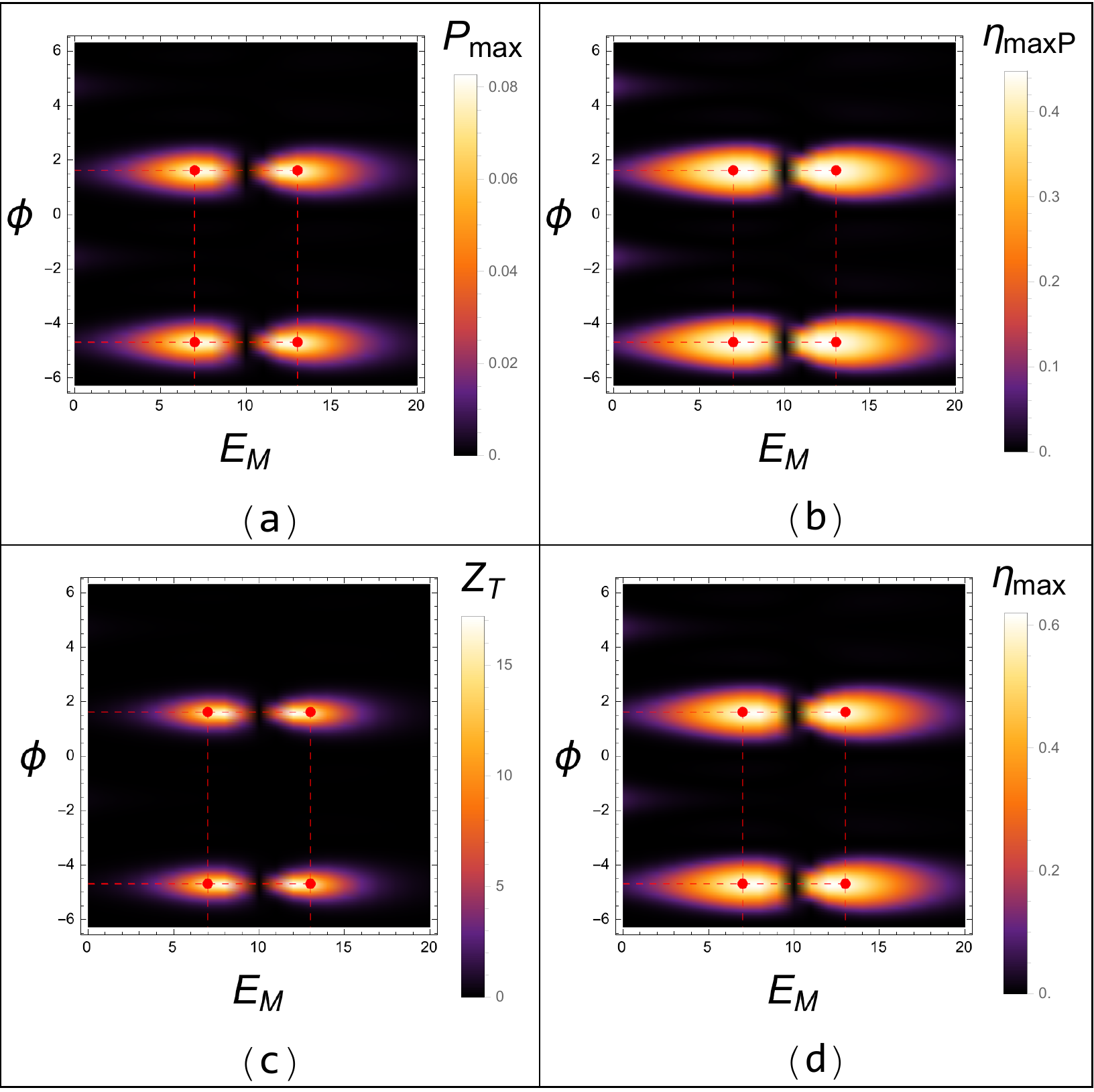}
\caption{Thermoelectric parameters for the Majorana quantum heat engine (a) Maximum power in units of $P_{0} = (k_{B}^{2}T \Delta \mathcal{T})/h $, (b) Efficiency at maximum power $\eta_{maxP}$ in units of $\eta_{c}$, {(c) the figure of merit $Z_T$}, and (d) maximum efficiency $\eta_{max}$ in units of $\eta_{c}$, vs. $\phi$ (in units of $\phi_{0}$) and $E_{M}$ (in $\mu eV)$ for $E_{F} = 10 \mu eV$, $\epsilon = 0.5$, $\Gamma = 20 \mu eV$, {with $\phi_0 = hc/e$}. The optimal regimes are highlighted with red points.}
\label{fig3}
\end{figure*}

\begin{figure*}[]
\center
\includegraphics[width=0.6\textwidth]{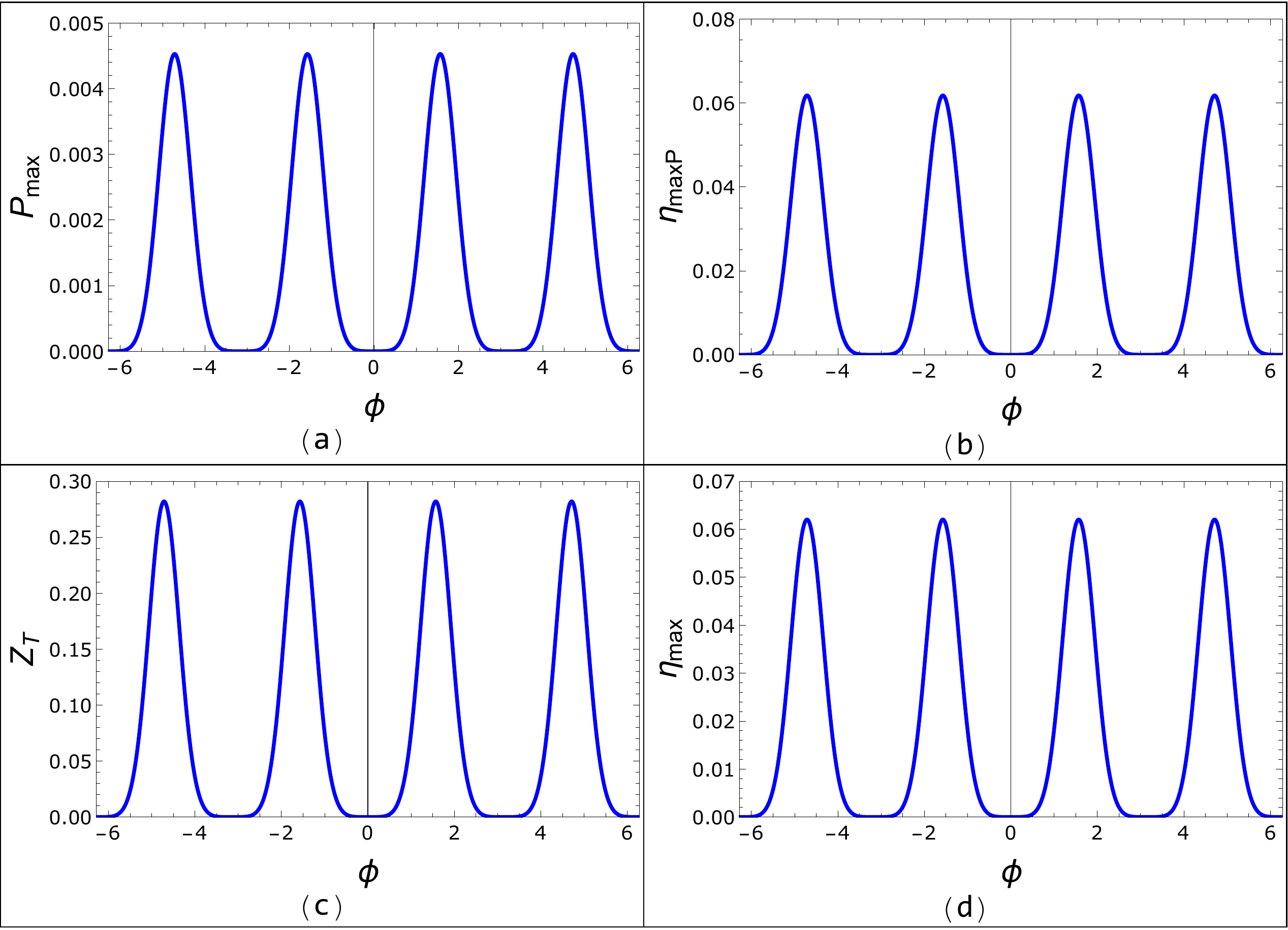}
\caption{Thermoelectric parameters for the Majorana quantum heat engine (a) Maximum power in units of $P_{0} = (k_{B}^{2}T \Delta \mathcal{T})/h $, (b) Efficiency at maximum power $\eta_{maxP}$ in units of $\eta_{c}$, {(c) the figure of merit $Z_T$}, and (d) maximum efficiency $\eta_{max}$ in units of $\eta_{c}$, vs. $\phi$ (in units of $\phi_{0}$) at $E_{M} = 0$ for $E_{F} = 10 \mu eV$, $\epsilon = 0.5$, $\Gamma = 20 \mu eV$, {with $\phi_0 = hc/e$}. The optimal regimes are highlighted with red points.}
\label{fig8}
\end{figure*}

\begin{equation}\label{eq1002}
\begin{split}
e^{\uparrow}_{i5} = e^{ik_{e}l_{1}}e^{\dfrac{-i\phi l_{1}}{L}} e^{\uparrow}_{iL},
e^{\uparrow}_{oL} = e^{ik_{e}l_{1}}e^{\dfrac{i\phi l_{1}}{L}} e^{\uparrow}_{o5}, \\
h^{\downarrow}_{i5} = e^{ik_{h}l_{1}}e^{\dfrac{i\phi l_{1}}{L}} h^{\downarrow}_{iL},
h^{\downarrow}_{oL} = e^{ik_{h}l_{1}}e^{\dfrac{-i\phi l_{1}}{L}} h^{\downarrow}_{o5},
\end{split}
\end{equation}
for the upper arm, right of STIM junction:
\begin{equation}\label{eq1003}
\begin{split}
e^{\uparrow}_{o6} = e^{ik_{e}l_{2}}e^{\dfrac{i\phi l_{2}}{L}} e^{\uparrow}_{oR},
e^{\uparrow}_{iR} = e^{ik_{e}l_{2}}e^{\dfrac{-i\phi l_{2}}{L}} e^{\uparrow}_{i6},\\
h^{\downarrow}_{o6} = e^{ik_{h}l_{2}}e^{\dfrac{-i\phi l_{2}}{L}} h^{\downarrow}_{oR},
h^{\downarrow}_{iR} = e^{ik_{h}l_{2}}e^{\dfrac{i\phi l_{2}}{L}} h^{\downarrow}_{i6},
\end{split}
\end{equation}
For the lower arm of the ring:
\begin{equation}\label{eq1004}
\begin{split}
e^{\uparrow}_{o3} = e^{ik_{e}l_{d}}e^{\dfrac{i\phi l_{d}}{L}} e^{\uparrow}_{o4},
e^{\uparrow}_{i4} = e^{ik_{d}l_{d}}e^{\dfrac{-i\phi l_{d}}{L}} e^{\uparrow}_{i3},\\
h^{\downarrow}_{o3} = e^{ik_{h}l_{d}}e^{\dfrac{-i\phi l_{d}}{L}} h^{\downarrow}_{o4},
h^{\downarrow}_{i4} = e^{ik_{h}l_{d}}e^{\dfrac{i\phi l_{d}}{L}} h^{\downarrow}_{i3},
\end{split}
\end{equation}
where $k_{e} = (E + E_{f})/\hbar v_{F}$, and $k_{h} = (E - E_{f})/\hbar v_{F}$. {Here, $L$ is the length of the entire Aharanov-Bohm ring with STIM junction in the upper arm of the AB ring, $l_1$ is the length of the upper arm to the left of the STIM junction, $l_2$ is the length of the ring of the upper arm to the right of STIM junction and $l_d$ is the length of the lower arm of the ring. We have a symmetric AB ring with length of upper and lower arm identical, i.e., $L = l_u + l_d$, with $l_u = l_d = 0.5L$ and $l_u = l_1 + l_2$ with $l_1 = l_2$, thus $l_1 = l_2 = 0.25 L$, where $L = \frac{\hbar v}{\Delta}$, where $v$ is the velocity of the electron or hole and $\Delta$ is the pairing potential of the superconductor present in the STIM junction. Furthermore, we have taken the same kinetic phases for the electrons/holes propagating in the inner and outer edge modes, regardless of their circumference. Electrons/Holes in an edge mode encounter zero scattering, thus we attribute similar phases to both.} $\phi$ is the Aharonov-Bohm flux taken in units of the flux quantum $\phi_{0} = hc/e$. Similarly, One can find the phase acquired by the spin-down electrons and holes, which is the exact mirror image of spin-up electrons and holes. Using Eqs. (\ref{eq1001}-\ref{eq1004}), we can calculate the scattering amplitudes and the resulting transmission probability. {The waves $h_{iL}^{\downarrow}$ and $e_{o1}^{\uparrow}$ are connected via the amplitude $s_{11}^{he, \downarrow \uparrow}$ for the spin-up electron to scatter from terminal 1 to itself as spin-down hole with transmission probability $T_{11}^{he, \downarrow \uparrow} = |s_{11}^{he, \downarrow \uparrow}|^2$. Similarly, the waves $e_{o2}^{\uparrow}$ and $e_{o1}^{\uparrow}$ are connected via the amplitude $s_{21}^{ee, \uparrow \uparrow}$ for the spin-up electron to scatter from terminal 1 to terminal 2 as a spin-up electron with the transmission probability $T_{21}^{ee, \uparrow \uparrow} = |s_{21}^{ee, \uparrow \uparrow}|^2$. Similarly, the waves $h_{o2}^{\downarrow}$ to $e_{o1}^{\uparrow}$ are connected via scattering amplitude $s_{21}^{he, \downarrow \uparrow}$ and transmission probability $T_{21}^{he, \downarrow \uparrow} = |s_{21}^{he, \downarrow \uparrow}|^2$.} We can then use the transmission probability $T(E,E_F)$ in Eq. (\ref{eq12}) to calculate the thermoelectric parameters to quantify the performance of our quantum heat engine/refrigerator. In the next section, we study the thermoelectric performance of our setup and outline the parameters where the setup works as a heat engine and where it works as a refrigerator.

\section{Thermoelectric Performance}\label{sec3}

\subsection{Majorana quantum heat engine}
In Fig. \ref{fig2}, we plot the maximum power, the efficiency at maximum power, maximum efficiency, and the figure of merit when the system is acting as a heat engine vs. the coupling between the MBS and the ring ($\Gamma_{1} = \Gamma_{2} = \Gamma$), and the coupling strength between individual MBS ($E_{M}$) for $E_{F} = 10 \mu eV$ and $\phi = \frac{\pi}{2} \phi_{0}$. {We choose $\phi = \frac{\pi}{2} \phi_{0}$ as this gives us the optimal performance (see Fig. \ref{fig3}).} Fig. \ref{fig2} (a) plots the maximum power output vs. $\Gamma$ and $E_{M}$. At $\Gamma = 0$, the STIM junction is wholly disconnected from the ring ($S_{MBS} = I$), and MBS are absent from the setup. For $\Gamma = 0$, we see that the maximum power is zero \cite{ramosweideman} as particle-hole symmetry (PHS) is preserved in the setup \cite{Leijnse_2014, proobing, ramosweideman}. As we increase $\Gamma$, the maximum power increases to a peak and slowly decays to zero. The maximum power is highest when $E_{M}$ is close to $E_{F}$ but not equal. If $E_{M} = E_{F}$, $P_{max} = 0$, i.e., if the Fermi energy of the TI is equal to the energy of coupling between the MBS, the thermoelectric response vanishes as the Seebeck coefficient is zero at precisely $E_{F} = \pm E_{M}$ \cite{ramosweideman, proobing}. The Seebeck coefficient changes sign at $E_{F} = \pm E_{M}$ as the dominant charge carrier changes from electrons to holes. As we move away from $E_{F}$ on both sides, the Seebeck coefficient increases in magnitude, thus improving the thermoelectric performance. On both sides of $E_{M}$, the maximum power increases to a peak and then decays to zero again. The rise and decay are rapid for lower $ \Gamma$, while the peaks are broadened for higher $\Gamma$. As we increase $\Gamma$, the STIM junction is strongly coupled to the ABI; thus, the breaking of PHS is more significant as we increase $\Gamma$ (Eq. (21)). This results in low electrical conductance and high Seebeck coefficients, leading to higher thermoelectric efficiency and lower maximum power (see Sec. V) \cite{helical444, Whitney}. When $\Gamma$ is zero, PHS is preserved, and efficiency and maximum power are zero.

In Fig. \ref{fig2} (b), we plot the efficiency at maximum power $\eta_{maxP}$ vs. $\Gamma$ and $E_{M}$. We see zero efficiencies at maximum power as expected for $\Gamma = 0$ (MBS absent). As we increase $\Gamma$, $\eta_{maxP}$ increases to a finite value and remains finite for higher $\Gamma$. Similar to the maximum power, since the Seebeck coefficient changes sign at exactly $E_{M} = \pm E_{F}$, we see that $\eta_{maxP}$ is highest when $E_{M}$ is close to $E_{F}$ and zero when $E_{M} = E_{F}$. As we disappear from $E_{F}$, $\eta_{maxP}$ increases rapidly and slowly decays to zero. The efficiency at maximum power, the figure of merit (Fig. \ref{fig2} (d)), and the maximum efficiency (Fig. \ref{fig2} (c)) behave similarly to each other {as we expect. However, the maximum power $P_{max}$ shows slightly different behaviour. While the figure of merit, and the maximum efficiency are higher for higher $\Gamma$, the maximum power is actually better for lower $\Gamma$. Thus, we observe a trade-off between the maximum power $P_{max}$, and the efficiency at maximum power {$\eta_{maxP}$} and the maximum efficiency $\eta_{max}$.} 
{From Fig. \ref{fig2} (c), the maximum efficiency can reach up to $0.65 \eta_{c}$ for $\Gamma = 30 \mu eV$ with $E_{M} = 8 \mu eV, 12 \mu eV$. The corresponding highest possible figure of merit for our setup is $Z_T = 20$ (Fig. \ref{fig2} (d)). The maximum power corresponding to these parameters is $P_{max} = 0.03 P_{0}$. However, at maximum efficiency, the output power is minimal \cite{Whitney}}. {As there is a trade-off between the maximum power and the efficiency at maximum power, we seek to identify the points of optimal performance. We define the points of optimal performance to be the points where the product of the maximum power and the efficiency at maximum power is the highest}. We look at Fig. \ref{fig2} (b), the efficiency at maximum power; we see that for $E_{M} = 8 \mu eV$, and $\Gamma = 20 \mu eV$, we have $P_{max} = 0.05 P_{0}$ and $\eta_{maxP} = 0.41\eta_{c}$, where $P_{0} = (k_{B}^{2}T \Delta \mathcal{T})/h$. The overall maximum efficiency $\eta_{max}$ at this point is $\eta_{max} = 0.45 \eta_{c}$. From Fig. \ref{fig3}, we see that for $\phi = \frac{\pi}{2} \phi_{0}$, and $E_{M} = 8 \mu eV$, we find $P_{max} = 0.05 P_{0}$, and $\eta_{maxP} = 0.41 \eta_{c}$. The overall maximum efficiency $\eta_{max}$ at this point is $0.4 \eta_{c}$. Thus, the optimal performance is for $\phi = \frac{\pi}{2} \phi_{0}$, $E_{M} = 8 \mu eV$, and $\Gamma = 20 \mu eV$ giving $P_{max} = 0.05 P_{0}$ and $\eta_{maxP} = 0.41 \eta_{c}$ {with the equivalent figure of merit $Z_T = 20$}. The overall maximum efficiency at the optimal regime is $\eta_{max} = 0.45 \eta_{c}$ with the corresponding figure of merit $Z_T = 20$. The optimal performance can also be reached at $\phi = \frac{\pi}{2} \phi_{0}$ and $E_{M} = 12 \mu eV$, with $\Gamma = 20 \mu eV$.
\par
In Fig. \ref{fig3}, we plot the same parameters vs. the Aharonov-Bohm flux $\phi$, and $E_{M}$ for $E_{F} = 10 \mu eV$, and $\Gamma = 20 \mu eV$. The maximum power, efficiency at maximum power, the figure of merit, and maximum efficiency are plotted against $\phi$ and $E_{M}$. The parameters are periodic with respect to $\phi$ with a $2 \pi$ period. For finite $E_{M}$, i.e., coupled MBS, the thermoelectric parameters are asymmetric with respect to $\phi$. The asymmetry in the thermoelectric parameters results from the breaking of time reversal symmetry and particle-hole symmetry by the MBS and the Aharonov-Bohm flux. The reasons for this asymmetry are detailed in Sec. V. Similar to Fig. \ref{fig2}, we see that the performance is best when $E_{M}$ is close to $E_{F}$, and zero when $E_{M} = E_{F}$. As we go away from $E_{F}$, the parameters rapidly increase to a peak and then decay to zero. We see that the thermoelectric parameters are periodic with respect to $\phi$ with a period of $2 \pi$. Thus, the optimal performance can also be reached for $\phi = -4.71 \phi_{0}$ while keeping $E_{M} = 8 \mu eV$ or $E_{M} = 12 \mu eV$ with $\Gamma = 20 \mu eV$.
\par
From Figs. \ref{fig2}-\ref{fig8}, we can study the behavior of the thermoelectric parameters with respect to the presence and nature of MBS in the system. Fig. \ref{fig2} shows that the thermoelectric parameters are zero when MBS is absent, i.e., $\Gamma = 0$. When MBS are present in the system, we can use the symmetry of the thermoelectric parameters with respect to the Aharonov-Bohm flux $\phi$ to distinguish their nature. From Fig. \ref{fig8}, we can see that for $E_{M} = 0$, or individual MBS, the thermoelectric parameters are symmetric with respect to $\phi$. For coupled MBS ($E_{M} \neq 0$), the parameters behave asymmetrically with respect to $\phi${, see Fig. \ref{fig3}}. Thus, the symmetry of the thermoelectric parameters can be used as a probe to detect the presence and nature of MBS in the setup. The reasons for this behavior are elaborated further in Sec. V.
\begin{figure}[]
\center
\includegraphics[width=0.4\textwidth]{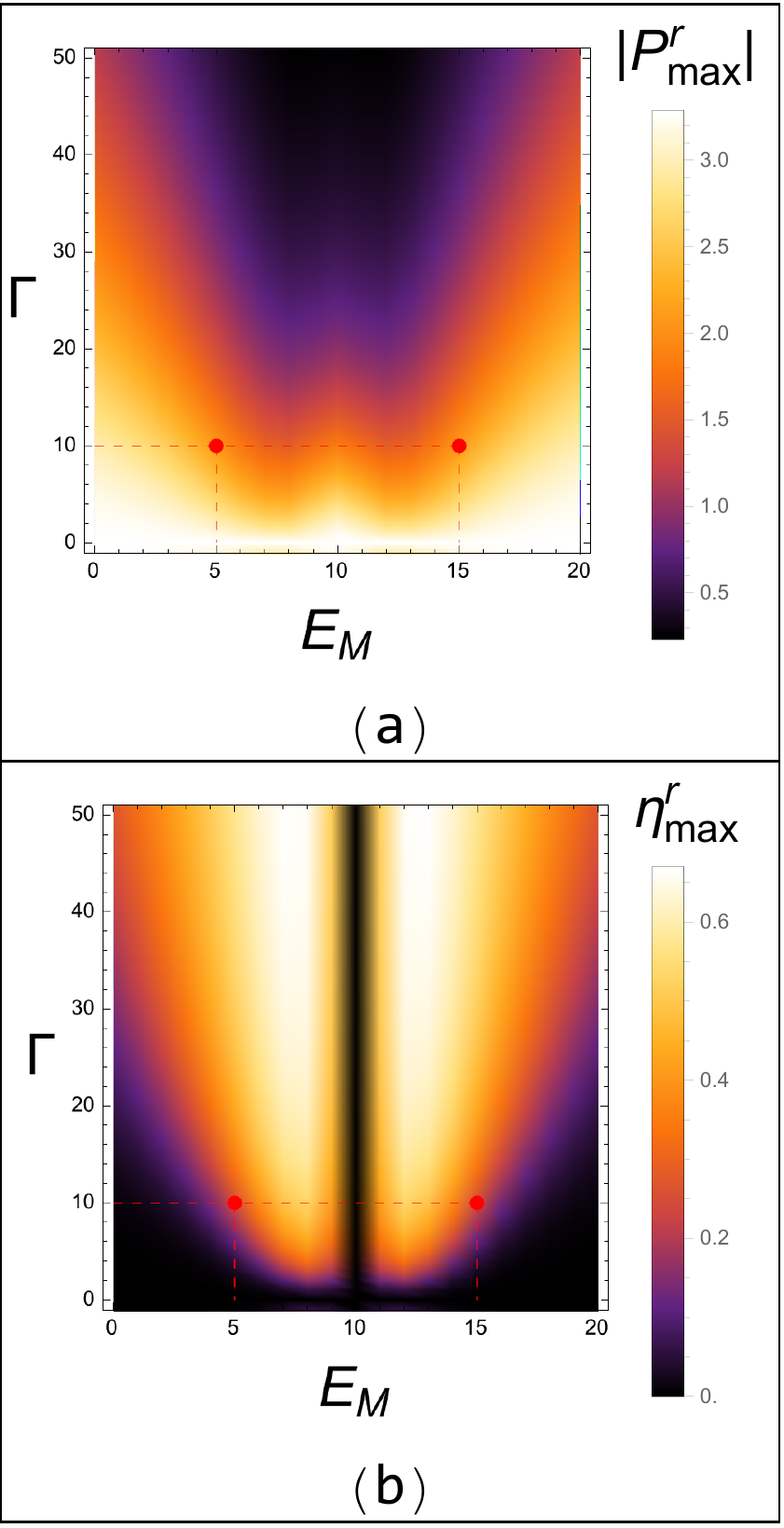}
\caption{(a)Magnitude of maximum cooling power $|P^{r}_{max}|$ in units of $P_{0}$, and (b) maximum COP $\eta^{r}_{max}$ in units of $\eta^{r}_{c}$ vs $\Gamma$ (in $\mu eV)$ and $E_{M}$ (in $\mu eV)$ for $E_{F} = 10 \mu eV$, $\epsilon = 0.5$, {$\phi = \frac{\pi}{2} \phi_{0}$}, {where $\phi_0 = hc/e$}. The optimal regimes are highlighted with red points.}
\label{fig4}
\end{figure}
\begin{figure}[]
\center
\includegraphics[width=0.4\textwidth]{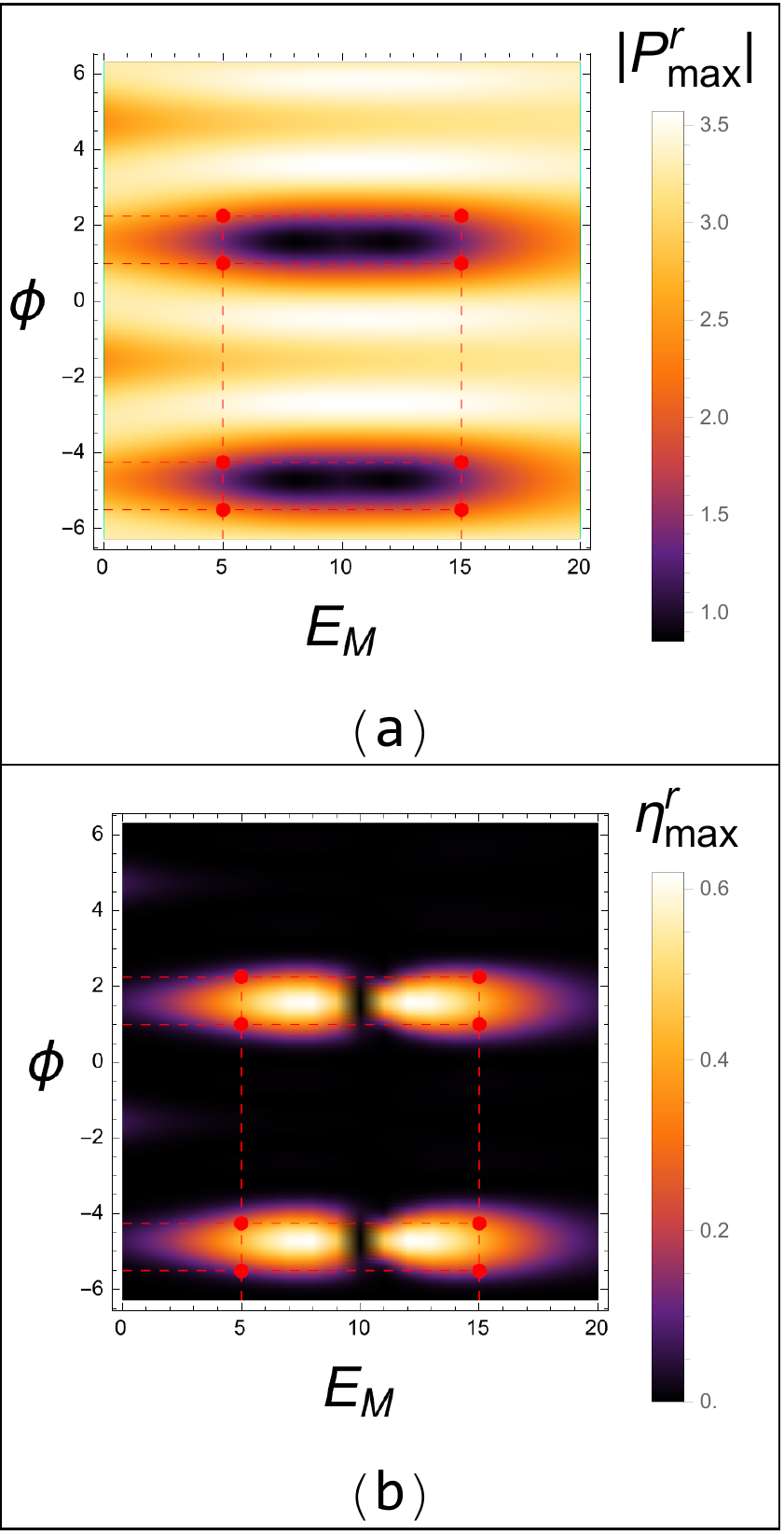}
\caption{(a) Magnitude of maximum cooling power $|P^{r}_{max}|$ in units of $P_{0}$, and (b) maximum COP $\eta^{r}_{max}$ in units of $\eta^{r}_{c}$, vs $\phi$ (in units of $\phi_{0}$) and $E_{M}$ (in $\mu eV)$ for $E_{F} = 10 \mu eV$, $\epsilon = 0.5$, {$\Gamma = 20 \mu eV$}, {where $\phi_0 = hc/e$}. The optimal regimes are highlighted with red points. }
\label{fig5}
\end{figure}

\begin{figure}[]
\center
\includegraphics[width=0.4\textwidth]{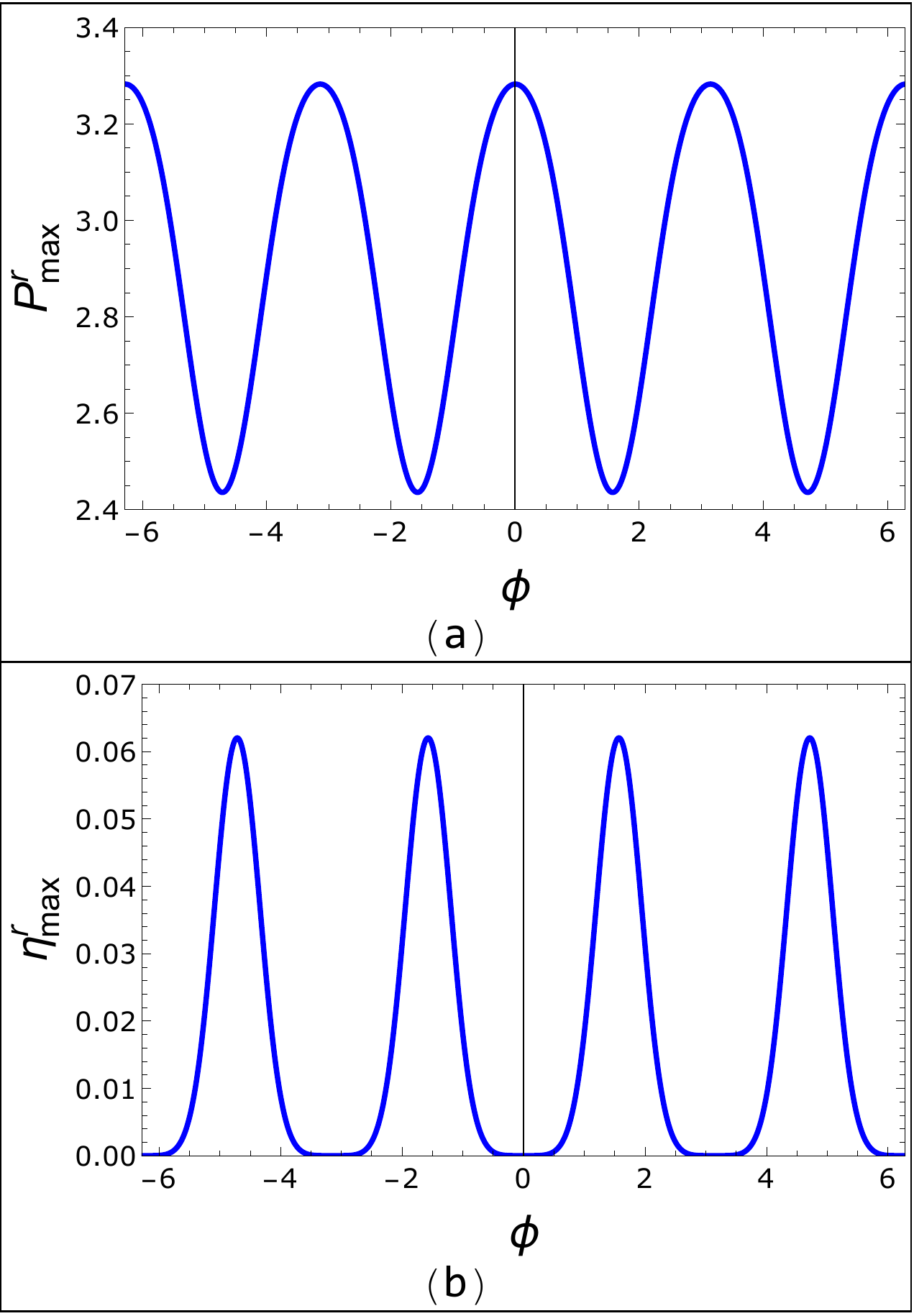}
\caption{(a) Magnitude of maximum cooling power $|P^{r}_{max}|$ in units of $P_{0}$, and (b) maximum COP $\eta^{r}_{max}$ in units of $\eta^{r}_{c}$, vs $\phi$ (in units of $\phi_{0}$) at $E_{M} = 0$ for $E_{F} = 10 \mu eV$, $\epsilon = 0.5$, {$\Gamma = 20 \mu eV$}, {where $\phi_0 = hc/e$}. The optimal regimes are highlighted with red points.}
\label{fig9}
\end{figure}

\subsection{Majorana quantum refrigerator}
This subsection looks at the setup's performance when acting as a quantum refrigerator. In {Fig. \ref{fig4} (a)}, we plot the cooling power at maximum efficiency vs. $\Gamma$ and $E_{M}$, with $E_{F} = 10 \mu eV$ and $\phi = \frac{\pi}{2} \phi_{0}$. The cooling power is maximum at very low $\Gamma$ and decays to zero as we increase $\Gamma$. In Fig. \ref{fig4} (b), we plot the maximum coefficient of performance vs. $\Gamma$ and $E_{M}$ for the same $E_{F}$ and $\phi$. Similar to Fig. \ref{fig2} (c), we see that for $\Gamma = 0$, i.e., when MBS are absent from the setup, the maximum COP is zero as particle-hole symmetry is preserved in the absence of MBS. We also see that for $E_{M} = E_{F}$, the maximum COP is zero. As we move away from $E_{F}$ in both directions, the maximum COP rapidly increases to a peak and then slowly decays. For higher $\Gamma$, since the electrical conductance is small and the Seebeck coefficient is high, we see a large COP and small cooling power. For lower $\Gamma$, the STIM junction is weakly coupled, and the electrical conductance is large. This results in lower COP but higher cooling power \cite{helical444, Whitney} (see Sec. V). {While the maximum COP $\eta^{r}_{max}$ behaves similar to the figure of merit $Z_T$, maximum cooling power $P^{r}_{max}$ behaves very differently}. Comparing Fig. \ref{fig4} (a) with Fig. \ref{fig4} (b), we see that the maximum COP and the cooling power are complementary to each other, i.e., as COP increases, cooling power decreases and vice-versa. {From Fig. \ref{fig4} (b), we can see that our setup can reach a maximum coefficient of performance (COP) $\eta^{r}_{max} = 0.65 \eta^{r}_{c}$ for $\Gamma = 30 \mu eV$, and $E_{M} = 8 \mu eV, 12 \mu eV$. The corresponding cooling power $P^{r}_{max} = 0.7 P_{0}$. Similarly, when the cooling power is maximum at {$P^{r}_{max} = 3.2$}, the corresponding COP is given by $\eta^{r}_{max} = 0.06$, which is very low. Since the transmission probability remains unchanged with respect to whether the setup performs as a quantum heat engine or as a quantum refrigerator, the figure of merit for the Majorana quantum refrigerator is the same as the figure of merit for the Majorana quantum heat engine for the same external parameters $E_{M}$, $E_{F}$, $\Gamma$, and $\phi$.} \\

\begin{figure*}
     \centering
     \begin{subfigure}[b]{0.45\textwidth}
         \centering
         \includegraphics[width=\textwidth]{fig6a.pdf}
         \caption{}
     \end{subfigure}
     \hspace{0.05cm}
     \begin{subfigure}[b]{0.45\textwidth}
         \centering
         \includegraphics[width=\textwidth]{fig6b.pdf}
         \caption{}
     \end{subfigure}
\caption{{Total transmission probability $T(E, E_F)$, versus electronic energy $E$ in units of $\mu eV$ for (a) maximal coupling ($\epsilon = 0.5$) and (b) non-maximal coupling ($\epsilon = 0.25$). The inset inside (a) is the Majorana absent case. Parameters in (a) and (b) are $E_F= 10 \mu eV$, $\phi = \frac{\pi}{2} \phi_0$ with $\phi_0 = hc/e$. For uncoupled and coupled Majorana case, we have $\Gamma = 51 \mu eV$.}}
\label{fig6}
\end{figure*}

\begin{figure*}[]
\center
\includegraphics[width=1.0\textwidth]{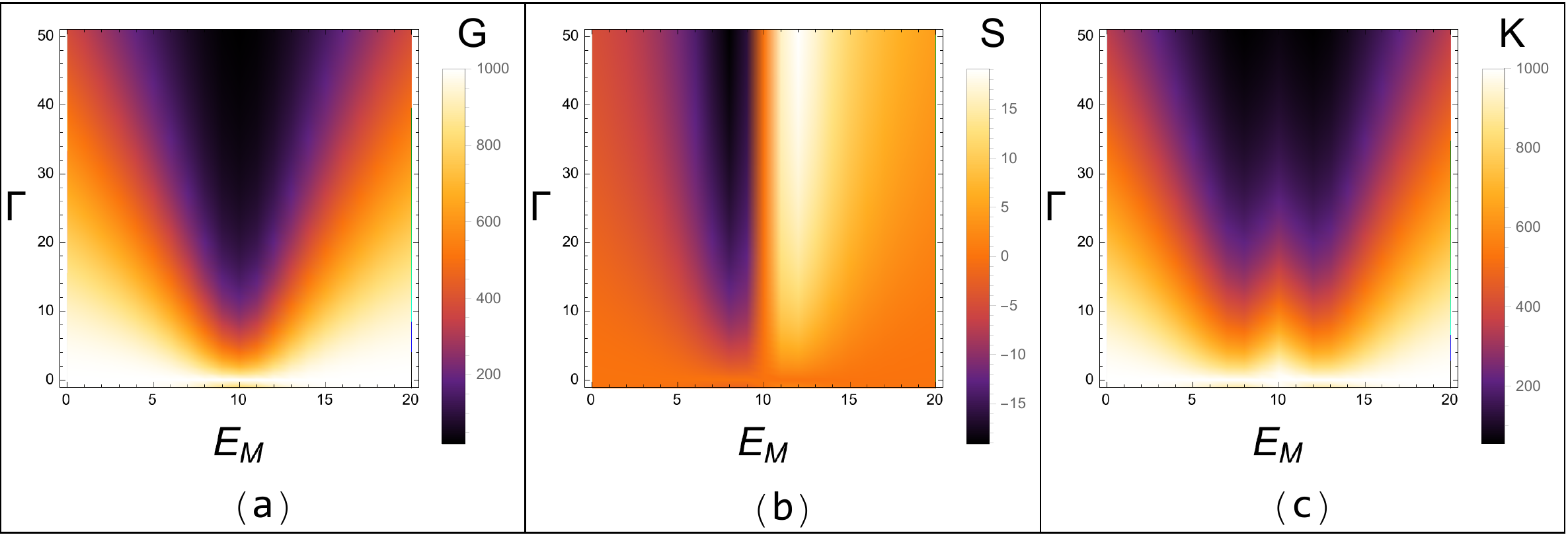}
\caption{{Thermoelectric transport parameters (a) Conductance ($G$) in units of $10^{-3} G_{0}$, (b) Seebeck coefficient in units of $\frac{k_B}{e}$, (c) Thermal conductance ($K$) in units of $10^{-3}\frac{\pi^2}{3} \frac{k_B^2 T}{h}$ vs. $\Gamma$ (in $\mu eV)$ and $E_{M}$ (in $\mu eV)$ for $E_{F} = 10 \mu eV$, $\epsilon = 0.5$, $\phi = \frac{\pi}{2} \phi_{0}$, with $\phi_{0} = hc/e$.}}
\label{fig7}
\end{figure*}

We plot the cooling power in Fig. \ref{fig5}(a) and maximum COP in Fig. \ref{fig5}(b) vs. $\phi$ and $E_{M}$ with $E_{F} = 10 \mu eV$, and {$\Gamma = 20 \mu eV$}. Like the heat engine parameters, the cooling power and the maximum COP are periodic with respect to $\phi$ with a period of $2 \pi$. From Fig. \ref{fig5}(a), we see that the cooling power is not necessarily high or low for $E_{M} = E_{F}$. Similar to Fig. {\ref{fig4}}, the cooling power and maximum COP are complementary. In regions with high COP, we see the lowest cooling power. By inspecting the figures, we can find the optimal balance of COP and cooling power with respect to $\phi$. {For the quantum refrigerator, we define the points of optimal performance to be the points where the product of the maximum COP {and $\eta_{max}^{r}$} and the corresponding maximum cooling power $P^{r}_{max}$ is the highest}. From Figs. \ref{fig4}-\ref{fig5}, we see that the optimal performance for setup as a quantum refrigerator is for $E_{F} = 10 \mu eV$, {$\Gamma = 20 \mu eV$}, with the optimal points $(E_{M}, \phi)$ given by $(5 \mu eV, \phi_{0})$, $(5 \mu eV, 2.14 \phi_{0})$, $(5 \mu eV, -4.14\phi_{0})$, $(5 \mu eV, -5.28\phi_{0})$, $(15 \mu eV, \phi_{0})$, $(15 \mu eV, 2.14 \phi_{0})$, $(15 \mu eV, -4.14\phi_{0})$, $(15 \mu eV, -5.28\phi_{0})$, we get $P^{r}_{max} = 1.8 P_{0}$ with $\eta^{r}_{max} = 0.45 \eta^{r}_{c}$ {and $Z_T = 20$}. We cannot plot COP at maximum cooling power, as in a quantum refrigerator, there is no bound on the maximum power \cite{Whitney}. Thus, we only plot the maximum COP and the cooling power at maximum COP.\\
Like the heat engine parameters, when MBS are absent ($\Gamma = 0$), PHS is unbroken, and the maximum COP is zero, but the cooling power remains constant. Since the maximum COP in the absence of MBS is zero, we cannot use the setup as a quantum refrigerator without MBS. In the presence of MBS, once again, we see that the cooling power and COP are symmetric with respect to $\phi$ in the presence of individual MBS ($E_{M} = 0$){, see Fig. \ref{fig9}} and asymmetric with respect to $\phi$ when coupled MBS are present ($E_{M} \neq 0$){, see Fig. \ref{fig5}}. Thus, the symmetry in cooling power and COP with respect to $\phi$ can also be used to probe the existence and nature of MBS.
\begin{table*}
\begin{tabular}{|c|c|c|c|}
\hline
Quantum heat Engine & \begin{tabular}[c]{@{}c@{}}Maximum power\\ (in units of $P_{0}$)\end{tabular} & \begin{tabular}[c]{@{}c@{}}Maximum Efficieny\\ (in units of $\eta_{c}$)\end{tabular} & \begin{tabular}[c]{@{}c@{}}Maximum possible \\ figure of merit \end{tabular} \\ \hline
Quantum Hall heat engine (two-terminal) \cite{engines} & 0.15 & 0.14 & 0.75 \\ \hline
Quantum Hall heat engine (three-terminal) \cite{PhysRevLett.114.146801} & 0.30 & 0.14 & 0.75 \\ \hline
Chaotic Cavity \cite{chaoticcavity} & 0.0066 & 0.01 & 0.04 \\ \hline
Strained graphene quantum heat engine \cite{strainedheat} & 0.268 & 0.4 & 4.44\\ \hline
{Aharanov-Bohm heat engine} \cite{PhysRevB.100.235442} & {NA} & {0.30} & {2.5}\\ \hline
{Normal-superconducting Aharanov-Bohm heat engine} \cite{blasi2022hybrid} & {NA} & {0.83} & {160}\\ \hline

Majorana quantum heat engine (This paper) & 0.09 & 0.65 & {20} \\ \hline
\end{tabular}
\caption{Comparing our MBS quantum heat engine to contemporary quantum heat engines. The values mentioned here are only in the linear response regime.}
\label{Table1}
\end{table*}
\begin{table*}[]
\resizebox{\textwidth}{!}{\begin{tabular}{|c|c|c|c|}
\hline
Quantum Refrigerator & \begin{tabular}[c]{@{}c@{}}Maximum cooling power\\ (in units of $P_{0}$)\end{tabular} & \begin{tabular}[c]{@{}c@{}}Maximum coefficient of performance\\ (in units of $\eta^{r}_{c})$\end{tabular} & \begin{tabular}[c]{@{}c@{}}Maximum possible\\ figure of merit\end{tabular} \\ \hline
Strained graphene quantum refrigerator \cite{strainedrefg} & 2.0 & 0.4 & 4.44 \\ \hline
Three-terminal quantum dot refrigerator \cite{threedot} & 0.87 & 0.4 & 4.44 \\ \hline
Magnon-driven quantum dot refrigerator \cite{magnon} & 0.8 & 0.2 & 1.25 \\ \hline

Majorana quantum refrigerator (This paper) & 3.3 & 0.65 & {20} \\ \hline
\end{tabular}}
\caption{Comparing our MBS quantum refrigerator to contemporary quantum refrigerators. The values mentioned here are only in the linear response regime.}
\label{Table2}
\end{table*}

\par
On comparing Figs. \ref{fig2}-\ref{fig3} with Figs. \ref{fig4}-\ref{fig5}, we see that the regions where the setup functions as a quantum heat engine are complementary to the regions where the setup functions as a quantum refrigerator, i.e., in the regions where the performance is high as a quantum heat engine, the performance as a quantum refrigerator is poor and vice-versa. Thus, the setup cannot simultaneously work as a quantum heat engine and a quantum refrigerator. The setup works as an optimal quantum heat engine or an optimal quantum refrigerator for any set of parameters. The Mathematica codes are available in GitHub \cite{s7}.

\section{Analysis}\label{sec4}
In Table \ref{Table1}, we compare the performance of the Majorana heat engine with other contemporary nanoscale quantum heat engines. We show that our proposed setup can outperform some quantum heat engines, like the cavity heat engine proposed in \cite{chaoticcavity}. We also show that our proposed setup can have higher efficiencies than other quantum heat engines like the three-terminal quantum spin Hall heat engine proposed in \cite{engines} with lower output power. The maximum power $P_{max}$ can reach up to $0.09 P_{0}$ for lower efficiency. {While the figure of merit $Z_T$ can reach up to 20 and the maximum efficiency $\eta_{max}$ can reach up to $0.65 \eta_{c}$ for our setup, at the highest maximum power $P_{max}$, $Z_T$ is equal to 20 and the corresponding efficiency at maximum power $\eta_{maxP} = 0.43 \eta_{c}$.}

In Table \ref{Table2}, we compare the performance of the Majorana refrigerator with other nanoscale quantum refrigerators. We see that our proposed setup outperforms the quantum refrigerators considered here. For the optimal parameters, the cooling power at maximum COP $P^{r}_{max} = 1.8 P_{0}$ and the corresponding maximum COP is given by $\eta^{r}_{max} = 0.45 \eta^{r}_{c}$. Fig. \ref{fig4} shows that the cooling power can reach up to {$3.2 P_{0}$} with very low COP. Conversely, from Fig. \ref{fig5}, we see that for very little cooling power, the maximum COP can go up to $0.65 \eta^{r}_{c}$ {and the corresponding highest figure of merit that can be achieved is $Z_T = 20$. At best performance, i.e. $P^{r}_{max} = 1.8 P_{0}$, the figure of merit $Z_T = 5.95$, which is very high compared to other quantum refrigerators.}

From Eq. (\ref{eq18}), we see that in the case of a quantum heat engine, the condition for a high $P_{max}$, within the linear response regime, is $S >> G$ \cite{helical444, strainedheat}, i.e., the Seebeck coefficient must be greater than the electrical conductance so that the thermoelectric response is high, also see Fig. \ref{fig7}. {The finite Seebeck coefficient arises only due to the Majorana bound states and the flux in our setup. When Majorana is absent ($E_M = 0, \Gamma = 0$), there is no electron-to-hole scattering, and only electron-to-electron scattering will occur. {In contrast, for $\Gamma = 0$ and with $\epsilon = 0.5$ (i.e., maximal coupling between ring and reservoirs) in the absence of Majorana, there can be multiple reflections, and the transmission probability is symmetric with respect to energy. There is direct electron/hole transmission from the left terminal to the right terminal with probability 1 at zero energy, and the transmission probability is energy-dependent and symmetric; see the inset of Fig. \ref{fig6}(a). In this situation, the particle-hole symmetry is not broken, and the Seebeck coefficient vanishes. Even if one takes asymmetric lengths of the ring, the transmission probability will still be symmetric with energy, and the particle-hole symmetry is not broken. One can consider an asymmetric ring where the upper arm length $l_u = l_1 + l_2 = 7L/12$, with $l_1 = L/4$ and $l_2 = L/3$, and the lower arm length $l_d = 5L/12$, ensuring that $l_u + l_d = L$. Even in this asymmetric case, the particle-hole symmetry remains intact, and the transmission probability continues to be symmetric with respect to energy, as in Fig. \ref{fig6}(a). However, to break the particle-hole symmetry with asymmetric lengths, one can apply gate voltages to the upper and lower arms of the ring, similar to the approach used in a normal metal Aharonov-Bohm ring as discussed in Refs.~\cite{PhysRevB.100.235442, Haack_2021}. With these gate voltages and asymmetric lengths, our setup can function as both a quantum heat engine and a quantum refrigerator, even in absence of MBS. With $\Gamma \ne 0$ (Majoranas present), the transmission probability becomes asymmetric with energy as a result of interference between different electron and hole paths through the upper and lower arms of the ring. Further, for $\Gamma = 0$, i.e., MBS are absent, but $\epsilon \ne 0.5$, there is back reflection from the coupler leading to interference (see Fig. \ref{fig6}(b)). In this case, the transmission probability is symmetric with energy and the Seebeck coefficient vanishes, which implies that the setup cannot work as a quantum heat engine or quantum refrigerator. Additionally, as shown in Refs.~\cite{PhysRevB.100.235442} and~\cite{blasi2022hybrid}, the asymmetric behavior of the transmission probability with respect to energy can occur by applying a gate voltage in the lower arm and having unequal lengths of the upper and lower arms of the AB ring.
 The Seebeck coefficient is not that high compared to the electrical conductance. Thus, we see a minimal maximum power when the setup acts as a quantum heat engine. On the other hand, in the case of a quantum refrigerator, for the cooling power $P_{max}^r$ to be high, the electrical conductance $G$ and the thermal conductance $K$ must be high. At the same time, the Seebeck coefficient must be low (Eq. (\ref{eq26})) \cite{helical444, strainedrefg}. {$K$ is proportional to $G$. Thus, it is clear that the maximum power for the quantum heat engine and the maximum cooling power in the quantum refrigerator are complementary as $S >> G$ for high $P_{max}$ in quantum heat engine and $S << G$ for high $P^{r}_{max}$ in quantum refrigerators, see Fig. \ref{fig7}.} This explains why the setup acts as a highly powerful quantum refrigerator but fails to act as an equally capable quantum heat engine. {We observe that at $\Gamma = 0$ and $\epsilon = 0.5$, a finite cooling power exists even in the presence of particle-hole symmetry. In this parameter regime, when no Majorana is present due to $\Gamma = 0$, and for $\epsilon = 0.5$, the maximal coupling case means the electron or hole transport is completely transparent via the couplers. The cooling power is defined as the heat current coming out of the cooler terminal (right terminal in our case), and it can be written as
$P_r = -J_2 =-( -L_{qV} \Delta V + L_{qT} \Delta \mathcal{T})$. In the presence of particle-hole symmetry, the thermoelectric coefficient $L_{qV}$ vanishes, but $L_{qT}$ remains finite. In this paper, we have considered $\epsilon = 0.5$. The transmission probability is symmetric with energy when Majorana is absent ($\Gamma = 0$) (see inset of Fig. \ref{fig6}(a)). For this regime, $L_{qV}$ vanishes, but $L_{qT}$ remains finite. Therefore, the cooling power is finite as $L_{qT} \ne 0$ and for a finite temperature bias $\Delta \mathcal{T}$. For $\Gamma \ne 0$ (Majoranas are present), particle-hole symmetry is broken as the transmission probability is energy-dependent (see Fig. \ref{fig6}(a)) and $L_{qV}$ is non-zero. One can also find finite cooling power at $\Gamma = 0$ and $\epsilon \ne 0.5$. For this case, the transmission probability due to interference is now energy-dependent (Fig. \ref{fig6}(b)) and is symmetric with energy. Therefore, $L_{qV}$ is zero for $\Gamma = 0$ for $\epsilon \ne 0.5$, and the cooling power is finite due to $L_{qT}$ and $\Delta \mathcal{T}$ only.} We also see that the performance is best when $E_{F}$ is close to $E_{M}$ but not equal. When $E_{M} = E_{F}$, there is a dip in the conductance, and the Seebeck coefficient vanishes because the dominant charge carriers change from electrons to holes; this results in zero efficiency and coefficient of performance. As we move away from $E_{F}$ on both sides, the conductance changes rapidly, and the Seebeck coefficient is very high, leading to high efficiency and coefficient of performance. {From Eq. (15), we see that the condition for a high figure of merit $Z_T$ is a large Seebeck coefficient $S$ and electrical conductance $\sigma$, coupled with a small thermal conductance $\kappa$. In our setup, both thermal conductance and electrical conductance can be small, but the high Seebeck coefficient $S$ makes up for the small electrical conductance as $Z_T$ is proportional to $S^{2}$ \cite{linearhigh, gatehigh}. This results in a large figure of merit when $E_{M}$ is close to $E_{F}$. The figure of merit $Z_T = 0$ when $E_{M} = E_{F}$.} Similar behavior of the thermoelectric figure of merit with respect to $E_{F}$ and $E_{M}$ has been reported in \cite{ramosweideman}, albeit with a massive reduction in magnitude. The increased thermoelectric efficiency {and the corresponding increased figure of merit for our setup is attributed to the presence of an Aharonov-Bohm flux, which breaks time-reversal symmetry (TRS), in addition to the MBS, which breaks particle-hole symmetry (PHS). The figure of merit in the presence of both MBS and an Aharonov-Bohm flux has not been studied before. While a large figure of merit such as $Z_T = 20$ is unusual, it is not unheard of. Ref. \cite{beta10} predicts that the figure of merit for mesoscopic thermoelectric devices can exceed $Z_T = 10$. Large figures of merit have been predicted before for quantum spin Hall insulators with $Z_T > 800$ \cite{linearhigh} and for three-dimensional TI nanowires with $Z_T > 120$ \cite{gatehigh}. A graphene-based quantum spin heat engine described in Ref. \cite{spinheateng} boasts a figure of merit as high as $Z_T = 50$. Ref. \cite{highZT1} sees similarly high values of the figure of merit with $Z_T = 100$.} {In Ref. \cite{Whitney}. Whitney derives upper bounds on the maximum power and maximum efficiency for a quantum heat engine operating in the non-linear response regime and the maximum cooling power and maximum COP for a quantum refrigerator operating in the non-linear response regime. Our quantum heat engine and quantum refrigerator operate in the linear response regime. Within the linear response regime, both the maximum efficiency and the maximum COP are bounded by the Carnot efficiency. From Eq. (1), we see that for a rapidly changing transmission function with very small magnitude, $S >> G$ \cite{Whitney, sothmannog}. Thus, we can deduce from Eq. (9) that $P_{max}$ can be very large with no upper bound in the linear response regime, but the corresponding efficiency will vanish \cite{Whitney, sothmannog}. Conversely, for a transmission function with a large magnitude and little change with respect to energy, $G, K >> S$ \cite{Whitney, sothmannog}. Thus, the maximum cooling power given by Eq. (18) can be very large with no upper bound in the linear response regime. However, at very high cooling power, the COP will vanish \cite{Whitney, sothmannog}.

We compare our quantities derived in the linear response regime with the non-linear upper bounds in Ref. \cite{Whitney} by Whitney. {The non-linear bound on the maximum efficiency for the quantum heat engine based on our setup is around $68 \%$ of the Carnot limit. Our quantum heat engine, operating in the linear response regime, can reach up to $65 \%$ of the Carnot limit with $\mathcal{T}_2= 11.5$mK and $\mathcal{T}_1 = 11.6$mK, which is lower than the non-linear upper bound, but close to it.} The non-linear bound on the maximum power of quantum heat engines, $P_{max}$, as calculated by Whitney, is $0.32 \frac{k_{B}^{2}T\Delta \mathcal{T}}{h}$. The maximum power of our quantum heat engine operating in the linear response regime can reach $0.09 \frac{k_{B}^{2}T\Delta \mathcal{T}}{h}$ which is almost {one-third} of the non-linear upper bound on maximum power. Our quantum refrigerator operating in the linear response regime can reach a maximum COP of up to $65 \%$ of the Carnot limit. The non-linear upper bound on the maximum cooling power derived by Whitney is given by,
\begin{equation}
    \mathcal{P}^{r}_{max} = \dfrac{N \pi^2}{12} P_{0},
\end{equation}
where $N$ is the total number of modes. For our quantum refrigerator with 8 modes, the non-linear upper bound on the maximum cooling power is $6.57 \frac{k_{B}^{2}T\Delta \mathcal{T}}{h}$. Our quantum refrigerator operating in the linear response can generate a cooling power of up to $3 \frac{k_{B}^{2}T\Delta \mathcal{T}}{h}$, which is around $45 \%$ of the upper bound for a non-linear quantum refrigerator.}

\par
Another application of our proposed setup is distinguishing the existence and nature of MBS in the setup. From Figs. \ref{fig2} and \ref{fig4}, we can see that when MBS is absent, i.e., $\Gamma = 0$, the thermoelectric parameters maximum power, maximum efficiency, the figure of merit, efficiency at maximum power, and maximum COP vanish, and the cooling power remains constant. The reason for the vanishing efficiencies is the preservation of TRS and PHS in the system in the absence of MBS. When MBS are present, particle-hole symmetry is broken, and we get finite thermoelectric parameters. MBS can be individual ($E_{M} = 0$) or coupled ($E_{M} \neq 0$). We use the symmetry with respect to the Aharonov-Bohm flux $\phi$ (see Figs. \ref{fig3} and \ref{fig5}) to distinguish between individual and coupled MBS. For individual MBS, we see that the thermoelectric parameters are symmetric with respect to $\phi$. For coupled MBS, the thermoelectric parameters are asymmetric with respect to $\phi$. The main reason for breaking the symmetry in the thermoelectric parameters is the breakdown of TRS due to the Aharonov-Bohm flux when MBS are coupled. When MBS are coupled ($E_{M} \neq 0$), the symmetry in the S-matrix given in Eq. (21) is broken as $y (E) \neq y (-E)$. It makes the S-matrix in Eq. (21) asymmetric, which results in asymmetric transport with respect to the reversal of Aharonov-Bohm flux due to the breakdown of TRS. If $E_{M} = 0$, only PHS is broken, and we get finite but symmetric thermoelectric parameters. This distinct behavior of the thermoelectric parameters for the absence of MBS, the presence of individual MBS, and the presence of coupled MBS can be used to detect the existence and nature of MBS in the setup \cite{proobing} via the maximum power and efficiencies of either the Majorana quantum heat engine or the Majorana quantum refrigerator.

\section{Conclusion}\label{sec5}
In this paper, we have proposed a setup that can act as a viable Majorana quantum heat engine and Majorana quantum refrigerator in separate parameter regimes due to the presence of Majorana-bound states. We have shown that the presence of Majorana bound states significantly increases the setup's performance as a quantum heat engine and a quantum refrigerator. We have outlined the effect of various parameters, such as the coupling strength of the STIM junction to the ring $\Gamma$, the strength of coupling between individual MBS $' E_{M}'$, and the Aharonov-Bohm flux $\phi$ piercing the ring. We show that the maximum efficiency and maximum COP of the system vanish in the absence of MBS. Thus, without MBS, we cannot use the setup as a quantum heat engine or quantum refrigerator. In the presence of MBS, we find that for the parameters $(E_{M}, \phi) = $ $(8 \mu eV, \frac{\pi}{2} \phi_{0})$, $(8 \mu eV, -4.71 \phi_{0})$, $(12 \mu eV, \frac{\pi}{2} \phi_{0})$ or $(12 \mu eV, -4.71 \phi_{0})$, with $\Gamma = 20 \mu eV$, the performance as a heat engine is optimal with $P_{max} = 0.05 P_{0}$, with $\eta_{max} = 0.45 \eta_{c}$, {and the effective figure of merit $Z_T = 20$}. For the parameters $\Gamma = 10 \mu eV$, with $(E_{M}, \phi)$ = $(5 \mu eV, \phi_{0})$, $(5 \mu eV, 2.14 \phi_{0})$, $(5 \mu eV, -4.14\phi_{0})$, $(5 \mu eV, -5.28\phi_{0})$, $(15 \mu eV, \phi_{0})$, $(15 \mu eV, 2.14 \phi_{0})$, $(15 \mu eV, -4.14\phi_{0})$, $(15 \mu eV, -5.28\phi_{0}),$ we see that the setup works as an optimal refrigerator with $P^{r}_{max} = 1.8 P_{0}$, and $\eta^{r}_{max} = 0.45 \eta_{c}$, {with the corresponding figure of merit $Z_T = 20$}, which is higher than some of the best contemporary nanoscale steady-state refrigerators. Since the efficiencies vanish in the absence of MBS, we have shown that it is the presence of MBS that makes it possible for the setup to work as a viable quantum heat engine and refrigerator.

When MBS are absent, the thermoelectric parameters maximum power, maximum efficiency, the figure of merit, efficiency at maximum power, and maximum COP vanish while the cooling power remains constant. When uncoupled MBS are present in the system, the thermoelectric parameters are symmetric with respect to $\phi$. When MBS are coupled, the thermoelectric parameters are asymmetric with respect to $\phi$. This behavior can be used to detect the presence and nature of MBS.

{Experimental realization of topological insulators with proximity-induced superconductivity is well-established, especially in the context of Majorana fermions \cite{Veldhorst2012, Sacepe2011}. Superconducting-ferromagnetic heterostructures have also been realized experimentally \cite{fs}. Thus, the proposed setup may be realized experimentally. We have ignored Joule heating in our work. Joule heating is given by $h_{Joule} = I^{2}RT$, where $I$ is the current flowing through the setup, $R$ is the resistance, and $T$ is the temperature. For ballistic transport, the resistance is very low. In addition, edge modes are not only ballistic but also topological, i.e., they suffer minimal scattering, much less than even ballistic modes. Thus, the resistance for transport via edge modes is very small. Thus, Joule heating for edge mode transport should be negligible. At low temperatures, i.e., $T \approx 0.1K$, phonon contribution is ignored because phonon contribution matters only at or near $T \approx 20K$ or more \cite{benettiphonon}. Thus, the contribution due to phonons is limited and can be neglected. The high performance and wide range of applicability can motivate researchers to work towards the experimental realization of our setup.}

{Furthermore, the entropy of the system can serve as a valuable tool for verifying whether the equilibrium state is governed by Majorana bound states. Recent studies have explored the intriguing distinction between equilibrium states dominated by Majorana bound states and those influenced by non-Majorana states \cite{PhysRevB.92.195312}. It has been established that equilibrium states governed purely by Majorana bound states exhibit a unique fractional entropy of \(\frac{1}{2} \ln(2)\). Any deviation from this characteristic entropy provides insight into the extent to which the physical quantities of the system are influenced by Majorana or non-Majorana bound states. In our current work, we do not yet quantify the relative contributions of Majorana and non-Majorana bound states to the physical quantities under investigation. However, we leave this as an open question for future research, inviting readers to explore this intriguing aspect further. This direction of inquiry is crucial, as it could enhance our understanding of the interplay between Majorana and non-Majorana states and their impact on measurable properties.
Notably, there has been growing experimental interest in computing entropy for equilibrium states influenced by non-Majorana states \cite{hartman2018direct, kleeorin2019measure}. These advancements underscore the potential of entropy measurements as a powerful diagnostic tool for identifying Majorana bound states. Developing robust experimental techniques to measure fractional entropy could significantly advance the detection and characterization of Majorana bound states, paving the way for deeper insights into their role in topological quantum systems. We believe this approach holds immense promise and could form the basis for transformative discoveries in the field.}

\appendix

\section{Derivation of charge and heat current in the quantum spin Hall ABI}\label{App1}
{The proposed setup is a two-terminal quantum spin Hall ABI. We explain the thermoelectric transport in our setup using Onsager relations \cite{sothmannog} and Landauer-Buttiker scattering theory \cite{BUTTIKER1983365}. According to the Landaeur-Buttiker formalism, the charge current in the terminal 1 due to the spin-up electron is (see, \cite{lambert1993multi, PhysRevB.53.16390}),}

{
\begin{equation}\label{eq1}
\begin{split}
    I_{1,e}^{\uparrow} = \frac{e}{h} \int_{-\infty}^{\infty}\! \! \! \! \! \! dE \bigg((1 - T_{1  1}^{e  e, \uparrow  \uparrow}) f_{1e}(E) + (-T_{1  1}^{e  h, \uparrow  \downarrow}) \\ f_{1h}(E)
    + (-T_{1  2}^{e  e, \uparrow  \uparrow}) f_{2e}(E) + (-T_{1  2}^{e  h, \uparrow  \downarrow})f_{2h}(E)\bigg),
    \end{split}
\end{equation}}

{where, $T_{1 \leftarrow i}^{e \leftarrow \alpha, \uparrow \leftarrow \sigma'}$ is the transmission probability for the particle of type $\alpha \in \{e, h \}$ with spin $\sigma' \in \{\uparrow, \downarrow\}$ to scatter from terminal $i \in \{1, 2 \}$ to left terminal terminal $1$ as an up-spin electron.} {$f_{1e} = \left(1 + e^{\frac{E - \mu_1}{k_B \mathcal{T}_1}}\right)^{-1}$ is the Fermi-Dirac distribution for electrons in left terminal 1, where $\mu_1 = E_F + e V_1$. Similarly, $f_{1h} = \left(1 + e^{\frac{-E + \mu_1}{k_B \mathcal{T}_1}}\right)^{-1}$ is the Fermi-Dirac distribution for holes in the left terminal 1. For the right terminal 2, $f_{2e} = \left(1 + e^{\frac{E - \mu_2}{k_B \mathcal{T}_2}}\right)^{-1}$ is the Fermi-Dirac distribution for electrons and $f_{2h} = \left(1 + e^{\frac{-E + \mu_2}{k_B \mathcal{T}_2}}\right)^{-1}$ is the same for holes, where $\mu_2 = E_F + e V_2$. Here, $V_1$ and $V_2$ are the voltage biases applied to the left and right terminals, and $\mathcal{T}_1$ and $\mathcal{T}_2$ are the temperatures of the left and right terminals. $E_F$ is the fermi energy of the reservoir. We consider $V_1 = -\Delta V$, $V_2 = 0$, $\mathcal{T}_1 = \mathcal{T} + \Delta \mathcal{T}$ and $\mathcal{T}_2 = \mathcal{T}$, where $\mathcal{T}$ is the equilibrium temperature of the setup. The electron and hole Fermi-Dirac distributions are related as $f_{1e} = 1 - f_{1h}$ and $f_{2e} = 1 - f_{2h}$.}  

{According to the Landaeur-Buttiker formalism, the charge current in the terminal 1 due to the spin-up electron is (see, \cite{lambert1993multi, PhysRevB.53.16390}),}

\begin{equation}\label{eq2}
\begin{split}
    {I_{1,e}^{\uparrow}} & {= \frac{e}{h} \int_{-\infty}^{\infty}\! \! \! \! \! \! dE \bigg(T_{2  1}^{e  e, \uparrow  \uparrow} (f_{1e}(E)-f_{2e}(E))  +  T_{2  1}^{e  h, \uparrow  \downarrow}} \\  & {(f_{1e}(E) - f_{2h}(E)) + T_{1 
 1}^{h  e, \downarrow  \uparrow}(f_{1e}(E) - f_{1h}(E))\bigg),}
 \end{split}
\end{equation}

{In the linear response regime, i.e., $e \Delta V \ll k_B \mathcal{T}$, the Taylor series expansion of $f_{1e}(E)$ around the Fermi energy $E_F$ and equilibrium temperature $\mathcal{T}$, i.e., when $\mu_1 \approx E_F$ and $\mathcal{T}_1 \approx \mathcal{T}$ is (see, \cite{steadystatebetter}),}

\begin{equation}\label{eq3}
     {f_{1e}(E) = f(E)  + \left. \frac{\partial f_{1e}}{\partial \mu_1} \right\vert_{\mu_i = \mu, \mathcal{T}_1 = \mathcal{T}}  e V_1 +  \left. \frac{\partial f_{1e}}{\partial \mathcal{T}_1} \right\vert_{\mu_i = \mu, \mathcal{T}_1 = \mathcal{T}}  \Delta \mathcal{T}.}
\end{equation}

{where $f(E) = \left(1 + e^{\frac{E - E_F}{k_B \mathcal{T}}}\right)^{-1}$ is the equilibrium Fermi-Dirac distribution at Fermi energy $E_F$ and equilibrium temperature $\mathcal{T}$ of the setup, which is also equal to $f_{2e}(E)$. Now, using the property of Fermi-Dirac distribution \cite{steadystatebetter},}

\begin{align} \label{eq200}
\begin{split}
{\left. \frac{\partial f_{1e}(E)}{\partial \mu_{1}}\right\vert_{\mu_1 = E_F, \mathcal{T}_1 = \mathcal{T}}} & = {\frac{-\partial f}{\partial E}}, \\
   {\left. \frac{\partial f_{1e}(E)}{\partial \mathcal{T}_{1}}\right\vert_{\mu_1 = E_F, \mathcal{T}_1 = \mathcal{T}}} & = {-\left(\frac{E - E_F}{k_B \mathcal{T}}\right)\frac{\partial f}{\partial E}},
   \end{split}
\end{align}
{which implies}
\begin{equation}\label{eq4}
     {f_{1e}(E) - f_{2e}(E) = -\frac{\partial f_{2e}}{\partial E} \left(e V_1 + \frac{E - E_F}{k_B \mathcal{T}}\Delta \mathcal{T}\right).}
\end{equation}

{Similarly, the Taylor series expansion of $f_{1e}(E) - f_{2h}(E)$ is given as}
\begin{align}\label{eq5}
\begin{split}
    {f_{1e}(E) - f_{2h}(E)} & {=  f_{2e}(E)  -  \frac{\partial f_{2e}}{\partial E} (e V_1 + \frac{E - E_F}{k_B \mathcal{T}}\Delta \mathcal{T})} \\& {- (1 -  f_{2e}(E_F))}
   \\ & {=  (2 f_{2e} - 1)  -   \frac{\partial f_{2e}}{\partial E}  (e V_1 + \frac{E - E_F}{k_B \mathcal{T}}\Delta \mathcal{T})}
    \end{split}
\end{align}

 {Now, the first term in the above expansion in Eq. (\ref{eq5}) makes zero contribution to transport since $f_{2e}$ is an equilibrium Fermi-Dirac distribution of the setup. Without any bias, there will be no charge and heat current and therefore the first term is ignored as it is an equilibrium contribution and is independent of both voltage and temperature biases. Therefore, the final expression of $f_{1e}(E) - f_{2h}(E)$ after the Taylor series expansion is,} 
\begin{equation}\label{eq6}
\begin{split}
     {f_{1e}(E) - f_{2h}(E)} &  {= f_{1e}(E) - f_{2e} (E)} \\
    &  {= -\frac{\partial f_{2e}}{\partial E} (e V_1 + \frac{E - E_F}{k_B \mathcal{T}}\Delta \mathcal{T}).}
    \end{split}
\end{equation}

 {Similarly, the Taylor series expansion of $f_{1e}(E) - f_{1h}(E)$ is}

\begin{subequations}
\begin{align}
    {f_{1e}(E) - f_{1h}(E)} &= {2 f_{1e}(E) - 1 = 2 f_{2e}(E) - 1} \nonumber \\
    &\quad {- 2\frac{\partial f_{2e}}{\partial E} \left( e V_1 + \frac{E - E_F}{k_B \mathcal{T}} \Delta \mathcal{T} \right)} \label{eq7a}
\end{align}
\begin{align}
    {-2\frac{\partial f_{2e}}{\partial E} \left( e V_1 + \frac{E - E_F}{k_B \mathcal{T}} \Delta \mathcal{T} \right)} &= {2 \left( f_{1e}(E) - f_{2e}(E) \right)} \label{eq7b}
\end{align}
\end{subequations}

 {Here, too, the first term of Eq. (\ref{eq7a}) does not play any role in transport as it is an equilibrium contribution and thus is ignored. Substituting Eqs. (\ref{eq4}), (\ref{eq5}) and (\ref{eq7b}) in Eq. (\ref{eq2}), we get}
 {
\begin{equation}\label{eq8}
    I_{1,e}^{\uparrow} = \frac{e}{h} \int_{-\infty}^{\infty}\! \! \! \! \! \! dE \bigg(2T_{1  1}^{h  e, \downarrow  \uparrow} +  T_{2  1}^{e  e, \uparrow  \uparrow} + T_{2  1}^{e  h, \uparrow  \downarrow}\bigg) (f_{1e}(E)-f_{2e}(E)),
\end{equation}}
{{Similarly, the down-spin hole current in terminal 1 is defined as (see, \cite{PhysRevB.53.16390, lambert1993multi})}}

{
\begin{equation}\label{eq9}
\begin{split}
    {I_{1,h}^{\downarrow} = \frac{-e}{h} \int_{-\infty}^{\infty}\! \! \! \! \! \! dE \, \, 
   \, \bigg((1 - T_{1  1}^{h  h, \downarrow  \downarrow}) f_{1h}(E) + (-T_{1  1}^{h  e, \downarrow  \uparrow})}\\ {f_{1e}(E) + (-T_{1  2}^{h  h, \downarrow   \downarrow}) f_{2h}(E) + (-T_{1  2}^{h  e, \downarrow  \uparrow})f_{2e}(E)\bigg),}
   \end{split}
\end{equation}}

{Here, too using the probability conservation $T_{1  1}^{h  h, \downarrow  \downarrow} + T_{1  1}^{e  h, \uparrow  \downarrow} + T_{2 
 1}^{h  h, \downarrow  \downarrow} + T_{2  1}^{e  h, \uparrow  \downarrow} = T_{1  1}^{h  h, \downarrow  \downarrow} + T_{1  1}^{e  h, \uparrow  \downarrow} + T_{1  2}^{h  h, \downarrow  \downarrow} + T_{1  2}^{h  e, \downarrow  \uparrow} = 1$ and following the same procedure used to derive {$I_{1,e}^{\uparrow}$}, one can simplify Eq. (10) to get,}

{
\begin{equation}\label{eq10}
\begin{split}
    {I_{1,h}^{\downarrow} = \frac{e}{h} \int_{-\infty}^{\infty}\! \! \! \! \! \! dE \, \, 
   \, \bigg(2T_{1  1}^{e  h, \uparrow  \downarrow} +  T_{2  1}^{h  h, \downarrow  \downarrow} + T_{2  1}^{e  h, \uparrow  \downarrow}  \bigg)} \\ {(f_{1e} - f_{2e})}.
   \end{split}
\end{equation}}

{In our setup, see Fig. \ref{fig1}, the direction of propagation of the spin-up electron and spin-down hole are the same, and they suffer scattering due to the couplers and the STIM junction in a similar fashion, see Figs. \ref{fig1}(b) and (c). Therefore, $T_{11}^{eh, \uparrow \downarrow} = T_{1  1}^{h  e, \downarrow  \uparrow}$, $T_{2  1}^{h  h, \downarrow  \downarrow} = T_{2  1}^{e  e, \uparrow  \uparrow}$
and $T_{2  1}^{e  h, \uparrow  \downarrow} = T_{2  1}^{h  e, \downarrow  \uparrow}$, which gives $I_{1,e}^{\uparrow} = I_{1,h}^{\downarrow}$. Thus, the current due to the up-spin electron and down-spin hole are exactly identical, i.e., $I_{1,e}^{\uparrow} = I_{1,h}^{\downarrow}$, although the electron and hole are of opposite charge. Similarly, one can prove $I_{1,e}^{\downarrow} = I_{1,h}^{\uparrow}$.}

{Due to the mirror symmetry of the setup Fig. \ref{fig1}, the current due to spin-up electron and spin-down electron are same, i.e., $I_{1,e}^{\uparrow} = I_{1,e}^{\downarrow}$. Similarly, for the heat current too $J_{1,e}^{\uparrow} = J_{1,e}^{\downarrow}$. Thus, the total charge current in the setup is $I = I^{\uparrow} + I^{\downarrow} = I_{1,e}^{\uparrow} + I_{1,h}^{\uparrow} + I_{1,e}^{\downarrow} + I_{1,h}^{\downarrow} = 4 I_{1,e}^{\uparrow}$. Similarly, the total heat current in the setup is $J = J^{\uparrow} + J^{\downarrow} = J_{1,e}^{\uparrow} + J_{1,h}^{\uparrow} + J_{1,e}^{\downarrow} + J_{1,h}^{\downarrow} = 4 J_{1,e}^{\uparrow}$. The total charge and heat current in the setup is then,}

{
\begin{equation}\label{eq11}
\begin{split}
    {I} & {= 4 I_{1,e}^{\uparrow} = \frac{4e}{h} \int_{-\infty}^{\infty} \! \! \! \! \! \! dE \, \, \, T(E,E_F)(f_{1e} - f_{2e})},\quad {\text{and}} \\ 
    {J} & {= 4 J_{1,e}^{\uparrow} = \frac{4}{h} \int_{-\infty}^{\infty} \! \! \! \! \! \! dE \, \, 
   \, (E - E_F + e \Delta V) \, \,  T(E,E_F) \, \, (f_{1e} - f_{2e})}
    \end{split}
\end{equation}}

{where, $T (E, E_F) = \left(2 T_{1  1}^{h  e, \downarrow  \uparrow} + T_{2  1}^{e  e, \uparrow  \uparrow} + T_{2  1}^{h  e, \downarrow  \uparrow} \right)$, and $T_{2  1}^{e  e, \uparrow   \uparrow}$ and $T_{2  1}^{h  e, \downarrow  \uparrow}$ are the transmission probabilities for non-local electron cotunneling and crossed Andreev reflection.}

\bibliography{apssamp}
\end{document}